\documentclass{ejm}

\usepackage{amsmath}
\usepackage{afterpage}
\usepackage[dvips]{graphicx}
\usepackage[normal]{subfigure}
\usepackage{amssymb}
\usepackage{subfig}
\def\ni{\noindent}
\def\proof{{\ni \bf \underline{Proof:} }}

\newcommand{\bsub}{\begin{subequations}}
\newcommand{\esub}{\end{subequations}$\!$}

\newcommand{\zt}{\tilde{z}}
\newcommand{\lt}{\tilde{l}}

\newcommand{\R}{{\mathbb{R}}}
\newcommand{\eps}{{\displaystyle \varepsilon}}
\newcommand{\inte}{{\int_{-1/\eps}^{1/\eps}}}
\newcommand{\lam}{{\lambda}}
\newcommand{\alp}{{\alpha}}
\newcommand{\kap}{{\kappa}}

\newcommand{\p}{\prime}

\newcommand{\sech}{\operatorname{sech}}

\renewcommand{\theequation}{\arabic{section}.\arabic{equation}}

\newtheorem{example}{Example}[section]

\newcommand{\bex}{\begin{example}\rm}
\newcommand{\eex}{\end{example}}

\renewcommand{\theequation}{\arabic{section}.\arabic{equation}}
\renewcommand{\thesection}{\arabic{section}}

\title[The Stability of Steady-State Hot-Spot Patterns for a Reaction-Diffusion
Model of Urban Crime]{The Stability of Steady-State Hot-Spot Patterns for a 
Reaction-Diffusion Model of Urban Crime}

\author[T. Kolokolnikov, M. J. Ward, J. Wei ]{%
  T.\ns K\ls O\ls L\ls O\ls K\ls O\ls L\ls N\ls I\ls K\ls O\ls V, \ns
   M.\ns J. \ns W\ls A\ls R\ls D, \ns
  \and  J.\ns W\ls E\ls I 
}
\affiliation{Theodore Kolokolnikov; Department of Mathematics,
 Dalhousie University, Halifax, Nova Scotia, B3H 3J5, Canada,

 Michael Ward; Department of Mathematics,  University of British Columbia, 
 Vancouver, British Columbia, V6T 1Z2, Canada,

Juncheng Wei, Department of Mathematics, Chinese University of Hong Kong,
 Shatin, New Territories, Hong Kong.}

\date{December 22, 2011}
\pubyear{2011}
\volume{000}
\pagerange{\pageref{firstpage}--\pageref{lastpage}}

\begin{document}

\label{firstpage}
\maketitle

\baselineskip=12pt

\begin{abstract}
The existence and stability of localized patterns of criminal activity
are studied for the reaction-diffusion model of urban crime that was
introduced by Short et.~al.~[Math. Models. Meth. Appl. Sci.,
  {\bf 18}, Suppl. (2008), pp.~1249--1267]. Such patterns,
characterized by the concentration of criminal activity in localized
spatial regions, are referred to as hot-spot patterns and they occur
in a parameter regime far from the Turing point associated with the
bifurcation of spatially uniform solutions. Singular perturbation
techniques are used to construct steady-state hot-spot patterns in one
and two-dimensional spatial domains, and new types of nonlocal
eigenvalue problems are derived that determine the stability of these
hot-spot patterns to ${\mathcal O}(1)$ time-scale instabilities. From
an analysis of these nonlocal eigenvalue problems, a critical
threshold $K_c$ is determined such that a pattern consisting of $K$
hot-spots is unstable to a competition instability if $K>K_c$. This
instability, due to a positive real eigenvalue, triggers the collapse
of some of the hot-spots in the pattern. Furthermore, in contrast to
the well-known stability results for spike patterns of the
Gierer-Meinhardt reaction-diffusion model, it is shown for the crime
model that there is only a relatively narrow parameter range where
oscillatory instabilities in the hot-spot amplitudes occur.  Such an
instability, due to a Hopf bifurcation, is studied explicitly for a
single hot-spot in the shadow system limit, for which the diffusivity
of criminals is asymptotically large. Finally, the parameter regime
where localized hot-spots occur is compared with the parameter regime,
studied in previous works, where Turing instabilities from a spatially
uniform steady-state occur.

\end{abstract}


\baselineskip=16pt 

\setcounter{equation}{0}
\setcounter{section}{0}
\section{Introduction} \label{section:1}

Recently, Short {\it et.~al.} \cite{s_1, s_2, s_3} introduced an
agent-based model of urban crime that takes into account repeat or
near-repeat victimization. In dimensionless form, the continuum limit
of this agent-based model is the two-component reaction-diffusion PDE
system 
\bsub\label{1:main}
\begin{align}
A_t  & =\eps^{2} \Delta A -A+ P A+ \alpha \,, \qquad x\in \Omega\,; \qquad
  \partial_n A = 0 \,, \qquad x\in \partial\Omega \,,\\
\tau P_t  & =D \nabla \cdot \left(\nabla P- \frac{2P}{A} \nabla A\right)
 - P A+ \gamma-\alpha \,, \qquad x\in\Omega\,; \qquad
  \partial_n P = 0 \,, \qquad x\in \partial\Omega \,,
\end{align}
\esub where the positive constants $\eps^2$, $D$, $\alpha$, $\gamma$
and $\tau$, are all assumed to be spatially independent. In this
model, $P(x,t)$ represents the density of the criminals, $A(x,t)$
represents the \textquotedblleft attractiveness\textquotedblright\ of
the environment to burglary or other criminal activity, and the
chemotactic drift term $-2D\nabla \cdot\left( P\frac{\nabla
  A}{A}\right)$ represents the tendency of criminals to move towards
sites with a higher attractiveness. In addition, $\alpha$ is the
baseline attractiveness, while ${(\gamma-\alpha)/\tau}$ represents the
constant rate of re-introduction of criminals after a burglary. For
further details on the model see \cite{s_1}.

\begin{figure}[tb]
\begin{center}%
\[
\includegraphics[width=0.31\textwidth]{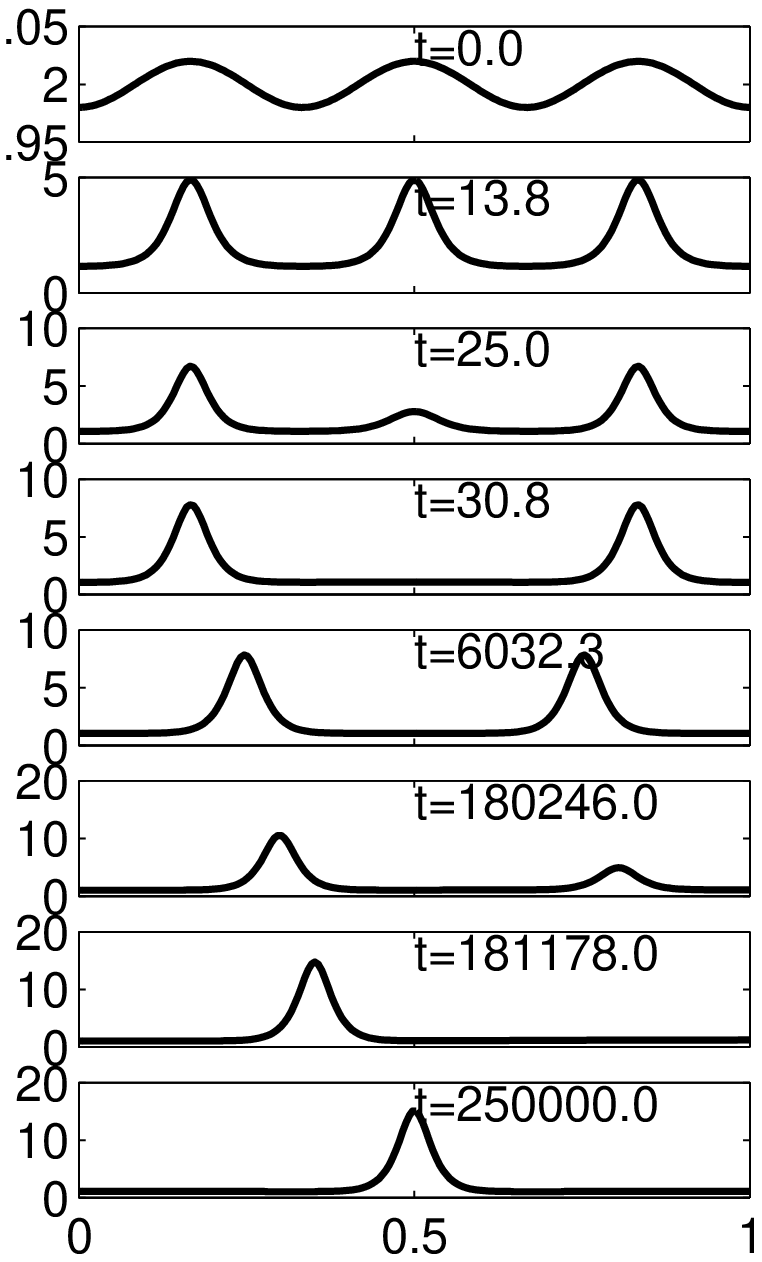}
\includegraphics[width=0.31\textwidth]{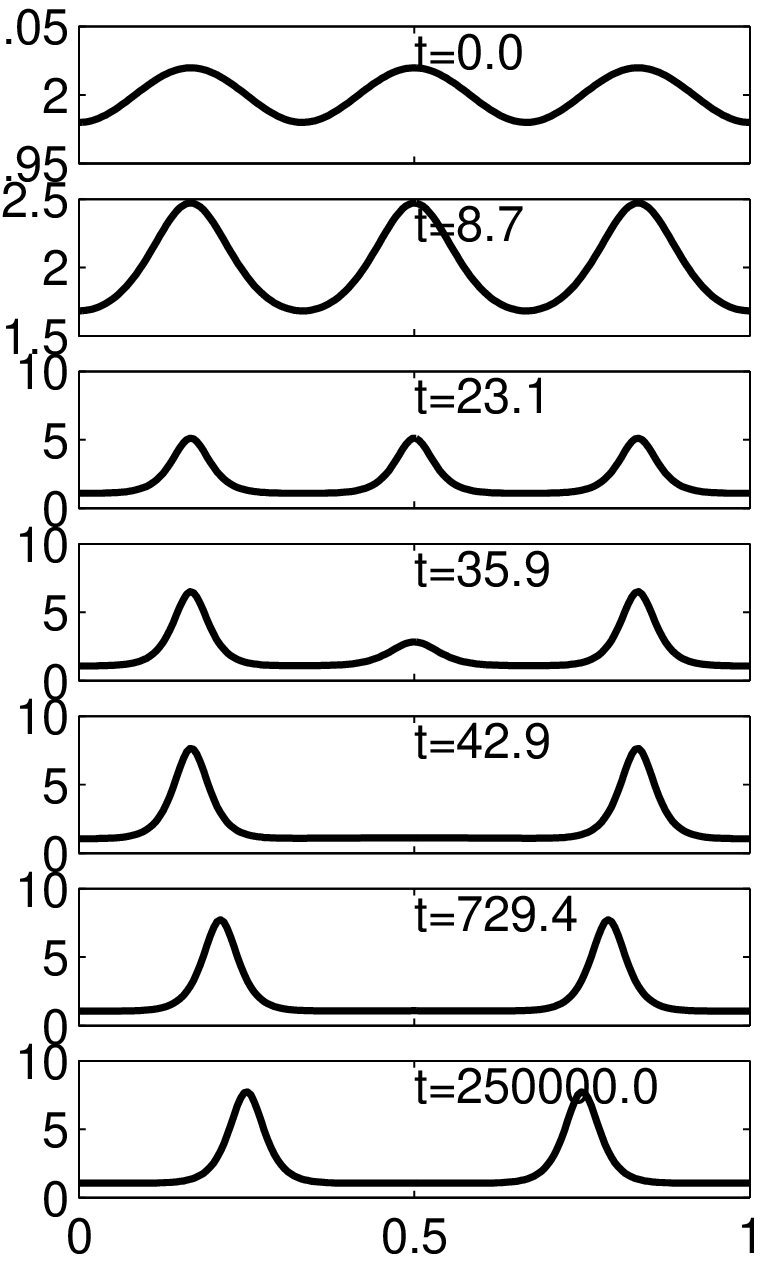}
\includegraphics[width=0.27\textwidth]{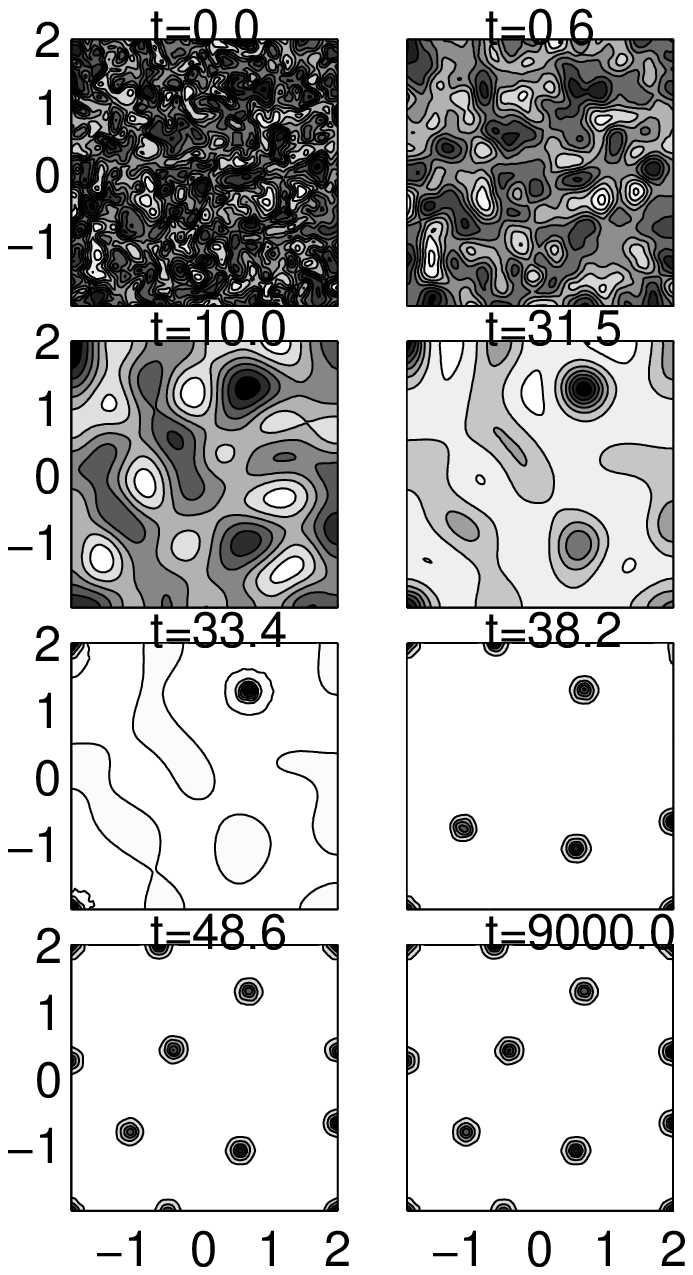}
\]%
\[
(a)~~~~~~~~~~~~~~~~~~~~~~~~~~~~~~~(b)~~~~~~~~~~~~~~~~~~~~~~~~~~~~~~~~~(c)
\]
\end{center}
\caption{Numerical solution of (\ref{1:main}) at different times, for initial
   data close to a spatially homogeneous steady-state. Plots of
  $A(x,t)$ are shown at the values of $t$ indicated. (a) One
  dimensional domain with parameter values $\alpha=1$, $\gamma=2$,
  $\varepsilon=0.02,$ $\tau=1$, $D=1.$ Initial conditions are
  $P(x,0)=1-\alpha/\gamma$ and $A(x,0)= \gamma(1-0.01\cos(6\pi x)).$
  Turing instability leads to a formation of three hot-spots; one is
  annihilated almost immediately due to a fast-time instability, while the
  second hot-spot is annihilated after a long time. (b) $D=0.5$ with all other
  parameters as in (a). Two hot-spots remain stable. (c) Numerical
  solution of (\ref{1:main}) in a two-dimensional square of width
  4. Parameters are $\alpha=1$, $\gamma=2$, $\varepsilon=0.08,$ $\tau=1$,
  $D=1.$ Initial conditions are $P(x,0)=1-\alpha/\gamma$ and $A(x,0)=
  \gamma(1+\text{rand}\ast0.001)$ where $\text{rand}$ generates a
  random number between 0 and 1.}%
\label{fig:turing}%
\end{figure}

In \cite{s_1}, the reaction-diffusion system (\ref{1:main}) with
chemotactic drift term was derived from a continuum limit of a
lattice-based model. It was then analyzed using linear stability
theory to determine a parameter range for the existence of a Turing
instability of the spatially uniform steady-state. A weakly nonlinear
theory, based on a multi-scale expansion valid near the Turing
bifurcation point, was developed in \cite{s_2, s_3} for (\ref{1:main})
for both one and two-dimensional domains. This theoretical framework
is very useful to explore the origins of various patterns that are
observed in full numerical solutions of the model. However, the major
drawback of a weakly nonlinear theory is that the parameters must be
tuned near the bifurcation point of the Turing instability. When the
parameters values are at an ${\mathcal O}(1)$ distance from the
bifurcation point, an instability of the spatially homogeneous
steady-state often leads to patterns consisting of \emph{localized
  structures}. Such localized patterns for the crime model
(\ref{1:main}), consisting of the concentration of criminal activity
in localized spatial regions, are referred to as either hot-spot or
spike-type patterns.  A localized hot-spot solution, not amenable to
an analytical description by a weakly nonlinear analysis, was observed
in the full numerical solutions of \cite{s_2}.

As an illustration of localization behavior, in
Fig.~\ref{fig:turing}(a) we plot the numerical solution to
(\ref{1:main}) in the one-dimensional domain $\Omega=\left[
  0,1\right]$ with parameter values $\alpha=1$, $\gamma=2$,
$\varepsilon=0.02$, $\tau=1$, and $D=1$. The initial conditions,
consisting of a small mode-three perturbation of the spatially
homogeneous steady-state $A_e=\gamma$ and $P_e=(\gamma-\alpha)/\gamma$
are first amplified due to linear instability. Shortly thereafter,
nonlinear effects become significant and the solution quickly becomes
localized leading to the formation of three hot-spots, as shown at
$t\approx14$. Subsequently, one of the hot-spots appears to be
unstable and is quickly annihilated. The remaining two hot-spots drift
towards each other over a long time, until finally around
$t\approx180,000$, another hot-spot is annihilated. The lone remaining
hot-spot then drifts towards the center of the domain where it then
remains.  Next, in Fig.~\ref{fig:turing}(b) we re-run the simulation
when $D$ is decreased to $D=0.5$ with all other parameters the same as
in Fig.~\ref{fig:turing}(a). For this value of $D$, we observe that
the final state consists of {\em two} hot-spots.  Similar complex
dynamics of hot-spots in a two-dimensional domain are shown in
Fig.~\ref{fig:turing}(c).

It is the goal of this paper to give a detailed study of the existence
and stability of steady-state localized hot-spot patterns for
(\ref{1:main}) in both one and two-dimensional domains in the singularly
perturbed limit
\begin{equation}
\varepsilon^{2}\ll D \,. \label{scalesep}%
\end{equation}
The assumption that $\varepsilon^{2}\ll D$ implies that the
length-scale associated with the change in the attractiveness of
potential burglary sites is much smaller than the length-scale over
which criminals explore new territory to commit crime. In this limit,
a singular perturbation methodology will be used to construct
steady-state hot-spot solutions and to derive new nonlocal eigenvalue
problems (NLEP's) governing the stability of these solutions.  From an
analysis of the spectrum of these NLEP's, explicit stability
thresholds in terms of $D$ and $\tau$ for the initiation of ${\mathcal
  O}(1)$ time-scale instabilities of these patterns are obtained. In a
one-dimensional domain, an additional stability threshold on $D$ for
the initiation of slow translational instabilities of the hot-spot
pattern is derived.  Among other results, we will be able to explain
both the fast and slow instabilities of the localized hot-spots
patterns as observed in Fig.~\ref{fig:turing}(a).

In related contexts, there is now a rather large literature on the
stability of spike-type patterns in two-component reaction-diffusion
systems with no drift terms. The theory was first developed in a
one-dimensional domain to analyze the stability of steady-state spike
patterns for the Gierer-Meinhardt model
(cf.~\cite{iww,dgk_0,mjww_1,mjww_3,ww_shad,vpd,ww_4,wei_rev}) and, in
a parallel development, the Gray-Scott model
(cf.~\cite{dgk_1,dgk_2,kww_1,mo_2,mo_3,kww_3,cw}).  The stability
theory for these two models was extended to two-dimensional domains in
\cite{ww_1,ww_2,ww_6,ww_5,ww_4,cw_1}.  Related studies for the
Schnakenburg model are given in \cite{iweiwin,mjww_2,ww_3}. The
dynamics of quasi-equilibrium spike patterns is studied for
one-dimensional domains in \cite{iw,dk,dkpr,swr,mckay}, and in a
multi-dimensional context in \cite{kmjw,kshad,kww_2,cw_1}.  More
recently, in \cite{kw} the stability of spikes was analyzed for a
reaction-diffusion model of species segregation with
cross-diffusion. A common feature in all of these studies, is that an
analysis of the spectrum of various classes of NLEP's is central for
determining the stability properties of localized patterns. A survey
of NLEP theory is given in \cite{wei_rev}, and in, a broader context,
a survey of phenomena and results for far-from-equilibrium patterns is
given in \cite{n}.

In contrast, for reaction-diffusion systems with chemotactic drift
terms, such as the crime model (\ref{1:main}), there are only a few
studies of the existence and stability of spike solutions. These
previous studies have focused mainly on variants of the well-known
Keller-Segel model (cf.~\cite{HP,kkw_0,PH,sww}).

We now summarize and illustrate our main results. In \S\ref{1:1d_s} we
construct a multi hot-spot steady state solution to (\ref{1:main}) on
a one-dimensional interval of length $S$. We refer to a symmetric
hot-spot steady-state solution as one for which the hot-spots are
equally spaced and, correspondingly, each hot-spot has the same
amplitude. In \S\ref{1:1d_as} asymmetric steady-state hot-spot
solutions, characterized by unevenly spaced hot-spots, are shown to
bifurcate from the symmetric branch of hot-spot solutions at a
critical value of $D$. 

In \S\ref{3:stab} we study the stability of steady-state $K$-hot-spot
solutions on an interval of length $S$ when $\tau={\mathcal O}(1)$.  A
singular perturbation approach is used to derive a NLEP that
determines the stability of these hot-spot patterns to ${\mathcal
  O}(1)$ time-scale instabilities. In contrast to the NLEP's arising
in the study of spike stability for the Gierer-Meinhardt model
(cf.~\cite{mjww_1}), this NLEP is explicitly solvable. In this way, a
critical threshold $K_{c+}$ is determined such that a pattern
consisting of $K$ hot-spots with $K>1$ is unstable to a competition
instability if and only if $K>K_{c+}$.  This instability, which
develops on an ${\mathcal O}(1)$ time scale as $\eps\to 0$, is due to
a positive real eigenvalue, and it triggers the collapse of some of
the hot-spots in the pattern. This critical threshold $K_{c+}>0$ is
the unique root of (see Principal Result 3.2 below)
\begin{equation}
  K \left( 1 + \cos\left({\pi/K}\right)\right)^{1/4} =
   \left( \frac{S}{2}\right) \left( \frac{2}{D}\right)^{1/4} 
   \frac{(\gamma-\alpha)^{3/4}}{\sqrt{\pi \eps \alpha}} \,. \label{Kc+}
\end{equation}
In addition, from the location of the bifurcation point associated with the
birth of an asymmetric hot-spot equilibrium, a further threshold $K_{c-}$
is derived that predicts that a $K$-hot-spot steady-state with $K>1$
is stable with respect to slow translational instabilities of the
hot-spot locations if and only if $K<K_{c-}$. This threshold is given
explicitly by (see (\ref{3:mstab_new}) below)
\begin{equation}
  K_{c-} =  \left( \frac{S}{2}\right) D^{-1/4} 
   \frac{(\gamma-\alpha)^{3/4}}{\sqrt{\pi \eps \alpha}} \,. \label{Kc}
\end{equation}
Since $K_{c-}<K_{c+}$, the stability properties of a $K$-hot-spot
steady-state solution with $K>1$ and $\tau={\mathcal O}(1)$ are as follows: 
stability when $K<K_{c-}$;
stability with respect to ${\mathcal O}(1)$ time-scale instabilities
but unstable with respect to slow translation instabilities when
$K_{c-}<K<K_{c+}$; a fast ${\mathcal O}(1)$ time-scale
instability dominates when $K>K_{c+}$.

As an illustration of these results consider again
Fig.~\ref{fig:turing}(a). From the parameter values in the figure
caption we compute from (\ref{Kc+}) and (\ref{Kc}) that $K_{c+}\approx
2.273$ and $K_{c-}\approx 1.995$. Therefore, we predict that the three
hot-spots that form at $t=13.8$ are unstable on an ${\mathcal O}(1)$
time-scale. This is confirmed by the numerical results shown at times
$t=25$ and $t=30.8$ in Fig.~\ref{fig:turing}(a).  We then predict from
the threshold $K_{c-}$ that the two-hot-spot solution will become
unstable on a very long time interval. This is also confirmed by the
full numerical solutions shown in Fig.~\ref{fig:turing}(a). In
contrast, if we decrease $D$ to $D=0.5$ as in Fig.~\ref{fig:turing}(b)
then we calculate from (\ref{Kc+}) and (\ref{Kc}) that $K_{c+}\approx
2.612$ and $K_{c-}\approx 2.372$. Our prediction is that the three
hot-spot solution that emerges from initial data will be unstable on
an ${\mathcal O}(1)$ time-scale, but that a two-hot-spot steady-state will
be stable. These predictions are again corroborated by the full
numerical results.

In \S\ref{4:hopf} we examine oscillatory instabilities of the
amplitudes of the hot-spots in terms of the bifurcation parameter
$\tau$ in (\ref{1:main}). From an analysis of a new NLEP with two
separate nonlocal terms, we show that an oscillatory instability of
the hot-spot amplitudes as a result of a Hopf bifurcation is not
possible on the regime $\tau\le {\mathcal O}(\eps^{-1})$. This
non-existence result for a Hopf bifurcation is in contrast to the
results obtained in \cite{mjww_1} for the Gierer-Meinhardt model
showing the existence of oscillatory instabilities of the spike
amplitudes in a rather wide parameter regime.  However, for the
asymptotically larger range of $\tau$ with $\tau={\mathcal
  O}(\eps^{-2})$, in \S\ref{4:hopf_g} we study oscillatory
instabilities of a single hot-spot in the simplified system
corresponding to letting $D\to \infty$ in (\ref{1:main}). In this
shadow system limit, we show for a domain of length one that low
frequency oscillations of the spot amplitude due to a Hopf bifurcation
will occur when $\tau>\tau_{c}$ where
\begin{equation*}
 \tau_{c} \sim 0.039759(\gamma-\alpha)^3\alpha^{-2} \eps^{-2} \,.
\end{equation*}

In \S\ref{5:multi} we extend our results to two dimensional
domains. We first construct a quasi-equilibrium multi hot-spot
pattern, and then derive an NLEP governing ${\mathcal O}(1)$
time-scale instabilities of the spot pattern. As in the analyses of
\cite{ww_1,ww_6, ww_2, ww_3, ww_4, ww_5} for the Gierer-Meinhardt and
Gray-Scott models, our existence and stability theory for localized
hot-spot solutions is accurate only to leading-order in powers of
${-1/\log\eps}$. In \S\ref{5:stab}, we show from
an analysis of a certain NLEP problem that for $\tau={\mathcal O}(1)$ a
$K$-hot-spot solution with $K>1$ is unstable when $K>K_{c}$ where
\begin{equation}
K_{c} \sim \frac{D^{-1/3} |\Omega| (\gamma-\alpha) \alpha^{-2/3} } 
 {\left[4\pi \left(\int_{\R^2} w^3 \, dy\right)^2\right]^{1/3} }
  \eps^{-4/3} \left(-\log\eps\right)^{1/3}
  \approx \left(0.07037\right)\,  D^{-1/3}
  \alpha^{-2/3} \left(\gamma-\alpha\right)  |\Omega| \varepsilon^{-4/3} 
\left( - \log{\varepsilon} \right)^{1/3} \,, \label{1:2d_th}
\end{equation}
as $\eps\to 0$.  Here $w$ is the radially symmetric ground-state
solution of $\Delta w - w + w^3=0$ in $\R^2$ and $|\Omega|$ is the
area of $\Omega$.  As an example, consider the parameter values as in
Fig.~\ref{fig:turing}(c), for which $|\Omega|=16$. Then, from
(\ref{1:2d_th}) we get $K_{c}\approx 44.48$. Starting with random
initial conditions, we observe from Fig.~\ref{fig:turing}(c) that at
$t=9000$ we have $K=7.5$ hot-spots, where we count boundary spots
having weight $1/2$ and corner spots having weight $1/4$. Since
$K<K_c$, this is in agreement with the stability theory.

Finally, in \S\ref{sec:discuss}, we contrast results for Turing
instabilities and Turing patterns with our results for localized
hot-spots. We also propose a few open problems.

\setcounter{equation}{0}
\setcounter{section}{1}
\section{Asymptotic Analysis of Steady-State Hot-Spot Solutions in 1-D}
\label{1:eq}

In the 1-D interval $x\in [-l,l]$, the reaction-diffusion system
(\ref{1:main}) is 
\bsub\label{2:1d_1}
\begin{align}
A_t  & =\eps^{2} A_{xx}-A+ P A+ \alpha \\
\tau P_t  & =D \left(  P_{x}- \frac{2P}{A} A_{x}\right)_{x}- P A+ \gamma-\alpha
 \,,
\end{align}
\esub
with Neumann boundary conditions $P_{x}(\pm l,t)=A_{x}(\pm l,t)=0$.
Since $P_x-\frac{2P}{A}A_x = (P/A^2)_x A^2$, it is convenient to introduce
the new variable $V$ defined by
\begin{equation}
   V = {P/A^2} \,,  \label{2:v_intro}
\end{equation}
so that (\ref{2:1d_1}) transforms to
\bsub\label{2:1d_2}
\begin{align}
A_t  & =\eps^{2}A_{xx}-A+ V A^{3}+ \alpha \,, \label{A}\\
\tau\left(A^2 V\right)_t & =D\left(  A^{2} V_{x}\right)_{x}- VA^{3}+
 \gamma -\alpha \,.\label{V}%
\end{align}
\esub

To motivate the $\eps$-dependent re-scaling of $V$ that facilitates
the analysis below, we suppose that $D \gg l^2$ and we integrate the
steady-state of (\ref{V}) over $-l<x<l$ to obtain that
$V={c/\int_{-l}^{l} A^3\, dx}$, where $c$ is some ${\mathcal O}(1)$
constant as $\eps\to 0$. Therefore, if $A={\mathcal O}(\eps^{-p})$ in
the inner hot-spot region of spatial extent ${\mathcal O}(\eps)$, we
conclude that $\int_{-l}^{l} A^3 \, dx = {\mathcal O}(\eps^{1-3p})$,
so that $V={\mathcal O}(\eps^{3p-1})$. In addition, from the
steady-state of (\ref{A}), we conclude that in the inner region near a
hot-spot centered at $x=x_0$ we must have $-A + A^3 V= {\mathcal
  O}(A_{yy})$, where $y={(x-x_0)/\eps}$ and $A={\mathcal
  O}(\eps^{-p})$. This implies that $-3p+(3p-1)=-p$, so that
$p=1$. Therefore, for $D\gg l^2$ we conclude that $V={\mathcal
  O}(\eps^2)$ globally on $-l<x<l$, while $A={\mathcal O}(\eps^{-1)}$
in the inner region near a hot-spot.  Finally, in the outer region we
must have $A={\mathcal O}(1)$, so that from the steady-state of
(\ref{V}), we conclude that $D \left(A^2 V_x\right)_x \sim \alpha
-\gamma ={\mathcal O}(1)$. Since $V={\mathcal O}(\eps^2)$, this
balance requires that $D={\mathcal O}(\eps^{-2})$. Since
$V={\mathcal O}(\eps^2)$ globally, while $A={\mathcal O}(\eps^{-1})$ in the
core of a hot-spot, we conclude that within a hot-spot of criminal
activity the density $P=VA^2$ of criminals is ${\mathcal O}(1)$.

In summary, this simple scaling analysis motivates the introduction of
new ${\mathcal O}(1)$ variables $v$ and $D_0$ defined by
\begin{equation}
  V = \eps^2 v \,, \qquad D = {D_0/\eps^2}  \,. \label{2:new}
\end{equation}
In terms of (\ref{2:new}), (\ref{2:1d_2}) transforms to
\bsub\label{2:1d}
\begin{align}
A_t & =\eps^{2}A_{xx}-A+ \eps^2 v A^{3}+ \alpha \,, \qquad -l<x<l \,; \qquad
 A_x(\pm l,t)=0 \,, \label{a}\\
\tau\eps^2 \left(A^2 v\right)_t & =D_0\left(  A^{2} v_{x}\right)_{x}-\eps^2
  vA^{3}+ \gamma -\alpha \,, \qquad -l<x<l \,; \qquad  v_x(\pm l,t)=0 \,.
\label{v}%
\end{align}
\esub

\subsection{A Single Steady-State Hot-Spot Solution}\label{1:1d_s}
We will now construct a steady-state hot-spot solution on the interval
$-l<x<l$ with a peak at the origin. In order to construct a
$K$-hot-spot pattern on a domain of length $S$, with evenly spaced
spots, we need only set $l={S/(2K)}$ and perform a periodic extension
of the results obtained below on the basic interval $-l<x<l$. As such,
the fundamental problem considered below is to asymptotically
construct a one-hot-spot steady-state solution on $-l<x<l$.

In the inner region, near the center of the hot-spot at $x=0$, we
expand $A$ and $v$ as
\begin{equation}
  A = \frac{A_0}{\eps} + A_1 + \cdots \,, \qquad v = v_0 + \eps v_1 + \eps^2
 v_2 + \eps^3 v_3 + \cdots
 \,,  \qquad y={x/\eps} \,.
\end{equation}
From (\ref{a}) we obtain, in terms of $y$, that $A_j(y)$ for $j=0,1$ satisfy
\bsub \label{2:isol}
\begin{align}
 A_{0}^{\p\p} - A_0 + v_0 A_0^3 &= 0 \,, \quad -\infty<y<\infty \,, 
  \label{2:isol_1}\\
 A_{1}^{\p\p} - A_1 + + 3 A_0^2 A_1 v_0 &= -\alpha - v_1 A_0^3 \,, \quad 
  -\infty<y<\infty \,.  \label{2:isol_2}
\end{align}
\esub
In contrast, from (\ref{v}), we obtain that $v_j$ for $j=0,1$ satisfy
\begin{equation}
 \left( A_0^2 v_0^\p\right)^\p=0 \,,  \qquad
 \left( A_0^2 v_1^\p + 2 A_0 A_1^\p v_0^\p \right)^\p=0 \,, \qquad 
  -\infty<y<\infty \,.
\end{equation}
In order to match to an outer solution, we require that $v_0$ and
$v_1$ are bounded as $|y|\to\infty$. In this way, we then obtain that
$v_0$ and $v_1$ must both be constants, independent of $y$.

We look for a solution to (\ref{2:isol}) for which the hot-spot has a 
maximum at $y=0$. The homoclinic solution to (\ref{2:isol_1}) with
$A_{0}^{\p}(0)=0$ is written as
\begin{equation}
  A_{0}(y) = v_{0}^{-1/2} w(y) \,, \label{2:ia0}
\end{equation}
where $w$ is the unique solution to the ground-state problem
\begin{equation}
  w^{\p\p} - w + w^3 = 0 \,, \quad -\infty<y<\infty \,; \quad w(0)>0 \,, 
\quad w^\p(0)=0 \,; \quad w\to 0 \quad \mbox{as} \quad |y|\to \infty \,,
\label{2:iw0}
\end{equation}
given explicitly by $w=\sqrt{2}\sech{y}$. Next, we decompose the
solution $A_1$ to (\ref{2:isol_2}) as
\begin{equation*}
  A_1 = \alp - \frac{v_1}{2v_0^{3/2}} w - 3\alpha w_1 \,,
\end{equation*}
where $w_1(y)$ satisfies
\begin{equation}
  L_{0} w_1 \equiv w_1^{\p\p} - w_1 + 3 w^2 w_1 = w^2 \,, \qquad
 -\infty<y<\infty  \,, \label{2:iw1}
\end{equation}
with $w_{1}^{\p}(0)=0$ and $w_{1}\to 0$ as $|y|\to \infty$.

A key property of the operator $L_0$, which relies on the
cubic exponent in (\ref{2:iw0}), is the remarkable identity that
\begin{equation}
   L_0 w^2 = 3 w^2 \,. \label{2:l0_iden}
\end{equation}
The proof of this identity is a straightforward manipulation of
(\ref{2:iw0}) and the operator $L_0$ in (\ref{2:iw1}). This property
plays an important role in an explicit analysis of the spectral
problem in \S \ref{3:stab}. Here this identity is used to provide an
explicit solution to (\ref{2:iw1}) in the form
\begin{equation*}
    w_1 = {w^2/3} \,. \label{2:w1sol}
\end{equation*}
In this way, in the inner region the two-term expansion for $A$ 
in terms of the unknown constants $v_0$ and $v_1$ is
\begin{equation}
  A(y) \sim \eps^{-1} A_0(y) + A_1(y) + \cdots  \,, \qquad A_0(y) =
 \eps^{-1} v_0^{-1/2} w(y) \,, \qquad A_1(y)=
  \alpha\left( 1 - [w(y)]^2\right)   - \frac{v_1}{2v_0^{3/2}} w(y)\,.
  \label{2:ia2t}
\end{equation} 

In the outer region, defined for $\eps\ll |x| \leq l$, we have that
$v={\mathcal O}(1)$ and that $A={\mathcal O}(1)$. From (\ref{2:1d}),
we obtain that
\begin{equation*}
   A = \alpha + o(1)  \,, \qquad v = h_0(x) + o(1) \,,
\end{equation*}
where from (\ref{v}), $h_0(x)$ satisfies
\begin{equation*}
   h_{0xx} = \zeta \equiv \frac{(\alpha - \gamma)}{D_0 \alpha^2} <0 \,,  \qquad
  0<|x|\leq l\,;  \qquad h_{0x}(\pm l)=0 \,,
\end{equation*}
subject to the matching condition that $h_0\to v_0$ as $x\to 0^{\pm}$. The
solution to this problem gives the outer expansion
\begin{equation}
  v \sim  h_{0}(x) = \frac{\zeta}{2} \left[ \left( l - |x|\right)^2 - l^2
\right] + v_0 \,, \qquad 0<|x|\leq l \,. \label{2:h0}
\end{equation}

Next, we must calculate the constants $v_0$ and $v_1$ appearing in
(\ref{2:ia2t}) and (\ref{2:h0}). We integrate (\ref{v}) over $-l<x<l$ and
use $v_x=0$ at $x=\pm l$ to get
\begin{equation*}
   \eps^2 \int_{-l}^{l} v A^3 \, dx = 2l (\alp-\gamma) \,. \label{2:glob}
\end{equation*}
Since $A={\mathcal O}(\eps^{-1})$ in the inner region, while 
$A={\mathcal O}(1)$ in the outer region, the dominant contribution to the 
integral in (\ref{2:glob}) arises from the inner region where
$x={\mathcal O}(\eps)$. If we use the inner expansion
$A=\eps^{-1}A_0+A_1 + o(1)$ from (\ref{2:ia2t}), and change variables 
to $y=\eps^{-1}x$, we obtain from (\ref{2:glob}) that
\begin{equation}
 v_0 \int_{-\infty}^{\infty} A_0^3 \, dy + \eps \left( 3v_0
\int_{-\infty}^{\infty} A_0^2 A_1 \, dy + v_1 \int_{-\infty}^{\infty} A_0^3
\right) + {\mathcal O}(\eps^2) = 2l (\alpha-\gamma) \,. \label{2:glob_exp}
\end{equation}
In (\ref{2:glob_exp}), we emphasize that the first two terms on the
left-hand side arise solely from the inner expansion, whereas the 
${\mathcal O}(\eps^2)$ term would be obtained from both the inner and outer
expansions. By equating coefficients of $\eps$ in (\ref{2:glob_exp}), we
obtain that
\begin{equation*}
  v_0 = 2l {(\gamma-\alp)/\int_{-\infty}^{\infty} A_0^3\, dy} \,, \qquad
  v_1 = -3v_0 { \int_{-\infty}^{\infty} A_0^2 A_1 \, dy/\int_{-\infty}^{\infty}
   A_0^3\, dy} \,, \qquad \label{2:v0v1_old}
\end{equation*}
Then, upon using (\ref{2:ia2t}) for $A_0$ and $A_1$, together with
$w=\sqrt{2}\sech{y}$, $\int_{-\infty}^{\infty} w^3\, dy=\sqrt{2}\pi$,
and $\int_{-\infty}^{\infty} w^4\, dy={16/3}$, we readily derive from
(\ref{2:v0v1_old}) that 
\begin{equation}
  v_0 = \frac{\pi^2}{2l^2 \left(\alpha-\gamma\right)^2} \,, 
 \qquad v_1 = 6 v_0^{3/2} \alp\left(
  \frac{\int_{-\infty}^{\infty} \left(w^2-w^4\right)\,
 dy}{ \int_{-\infty}^{\infty} w^3 \, dy} \right) = 
 -\frac{4\sqrt{2}}{\pi} v_{0}^{3/2} \alp
 = -\frac{2\alp\pi^2}{l^3 (\gamma-\alp)^3} \,. \label{2:v0v1}
\end{equation}

We summarize our result for a single steady-state hot-spot solution as follows:

\noindent {\bf \underline{Principal Result 2.1}}: {\em Let $\eps\to 0$,
and consider a one-hot-spot solution centered at the origin for (\ref{2:1d})
on the interval $|x|\leq l$. Then, in the inner region $y={x/\eps}=
{\mathcal O}(1)$, we have}
\begin{equation}
   A(y) = \frac{w}{\eps\sqrt{v_0}} + \alp\left( 1 + \frac{2\sqrt{2}}{\pi}
 w - w^2 \right) + o(1) \,, \qquad
 v\sim v_0 + \eps v_1 + \cdots \,. \label{2:avsum}
\end{equation}
{\em In addition, in the inner region, the leading-order steady-state
criminal density $P$ from (\ref{2:1d_1}) is $P\sim w^2$.
Here $w=w(y)=\sqrt{2}\sech{y}$ is the homoclinic of (\ref{2:iw0}), while
$v_0$ and $v_1$ are given in (\ref{2:v0v1}). In the outer region,
${\mathcal O}(\eps)<|x|\leq l$, then}
\begin{equation}
   A \sim \alpha + o(1) \,; \qquad v \sim \frac{\zeta}{2}
  \left( (l-|x|)^2 - l^2 \right) + v_0 + o(1) \,, \qquad
 \zeta \equiv \frac{(\alpha - \gamma)}{D_0 \alpha^2} <0 \,. \label{Avouter}
\end{equation}

\begin{figure}[tb]
\begin{center}%
$$
\includegraphics[width=0.4\textwidth]{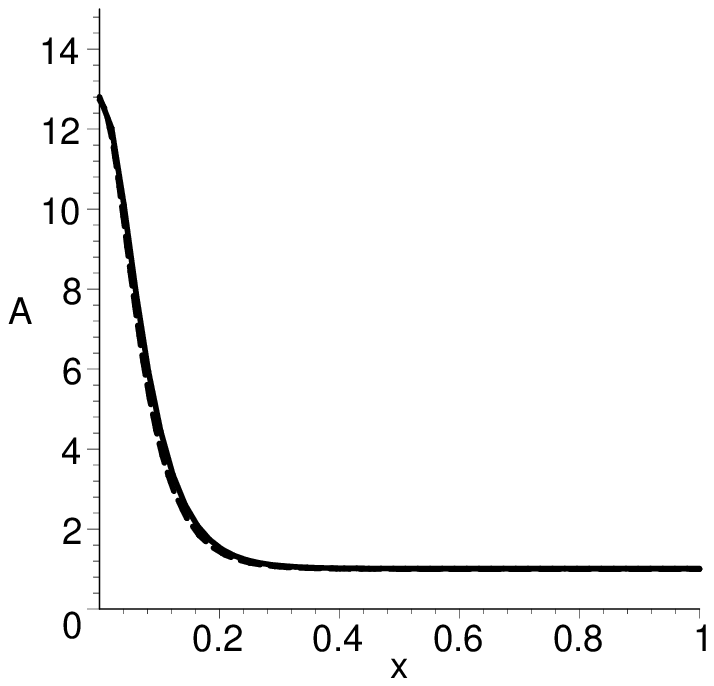}
~~~~~~~~~~~
\includegraphics[width=0.4\textwidth]{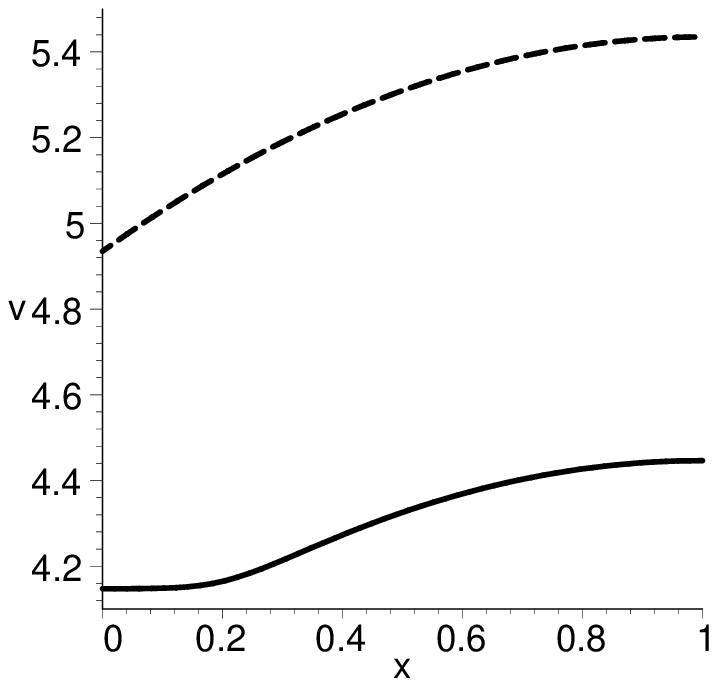}
$$
$$(a)~~~~~~~~~~~~~~~~~~~~~~~~~~~~~~~~~~~~~~~~~~~~~~~~~~~~~~~~~~~~(b)$$
\end{center}
\caption{Steady-state solution in one spatial dimension. Parameter values 
are $D_{0}=1,\varepsilon=0.05, \alpha=1, \gamma=2$, and $x\in[0,1].$ 
(a) The solid line is the steady state solution $A(x)$ of (\ref{2:1d}) 
computed by solving the boundary value problem
numerically. The dashed line corresponds to the first-order composite
approximation given by (\ref{Aunif}) (b) The solid line is the steady state
solution for $v(x).$ Note the ``flat knee'' region obtained from the full
numerical solution in the inner region near the center of the hot-spot.
The dashed line is the leading-order asymptotic result (\ref{Avouter}). }
\label{fig:ss1d} \end{figure}

Note that to get a solution for $A$ which is uniformly valid in both
inner and outer region, we can combine the formulas (\ref{2:avsum})
and (\ref{Avouter}).  The resulting first-order composite solution is
given explicitly by
\begin{equation}
A \sim \left(\frac{2l(\gamma -\alpha)}{\pi\eps}-\alpha  \right)
\sech\left(\frac{x}{\eps}\right)+\alpha \,.
\label{Aunif}
\end{equation}
For a specific parameter set, a comparison of the full numerical
steady-state solution of (\ref{2:1d}) with the composite asymptotic
solution (\ref{Aunif}) is shown in Fig.~\ref{fig:ss1d}. A comparison
of numerical and asymptotic values for $A(0)$ and $v(0)$ at various
$\eps$ is shown in Table~\ref{tab:equil}. From this table we note that
the two-term asymptotic expansion for $v(0)$ agrees very favorably with full
numerical results.

\begin{table}
\centering
\begin{tabular}{c|c|c|c|c|c|c}
\hline $\eps$ & $A(0)$ (num) & $A(0)$ (asy1) &
$A(0)$ (asy2)  & $v(0)$ (num) & $v(0)$ (asy1) &
$v(0)$ (asy2) \\ \hline
0.1    & 6.281 & 6.366 & 6.003   & 3.5844 & 4.935 & 2.961\\
0.05   &12.805 & 12.732 & 12.369 & 4.1474 & 4.935 & 3.948\\
0.025  &25.628 & 25.465 & 25.101 & 4.4993 & 4.935 & 4.441\\
0.0125 &51.145 & 50.930 & 50.566 & 4.7039 & 4.935 & 4.688 \\ \hline
\end{tabular}
\caption{Comparison of numerical and asymptotic results for the
  amplitude $A_\text{max}\equiv A(0)$ and for $v(0)$ of a one-hot-spot
  solution on $[-1,1]$ with $D_0=1$, $\gamma=1$, and $\alpha=2$. The
  1-term and 2-term asymptotic results for $A_\text{max}$ and 
  $v(0)$ are obtained from (\ref{2:avsum}).}
\label{tab:equil}
\end{table}

\subsection{Asymmetric Steady-State $K$-Hot-Spot Solutions}\label{1:1d_as}

In the limit $\eps \to 0$, we now construct an asymmetric steady-state
$K$-hot-spot solution to (\ref{2:1d}) in the form of a sequence of
hot-spots of different heights. This construction will be used to
characterize the stability of symmetric steady-state $K$-hot-spot
solutions with respect to the small eigenvalues $\lambda = o(1)$ in the 
spectrum of the linearization. Since the asymmetric solution
is shown to bifurcate from the symmetric branch, the point of the
bifurcation corresponds to a zero eigenvalue crossing along the
symmetric branch.  To determine this bifurcation point, we compute
$v(l)$ for the one-hot-spot steady-state solution to (\ref{2:1d}) on
$|x|\leq l$, where $l>0$ is a parameter. This canonical problem is
shown to have two different solutions. A $K$-hot-spot asymmetric solution
to (\ref{2:1d}) is then obtained by using translates of these two
local solutions in such a way to ensure that the resulting 
solution is $C^1$ continuous. Since the details of the construction of
the asymmetric solution is very similar to that in \cite{mjww_2} for
the Schnakenburg model, we will only give a brief outline of the
analysis. 

The key quantity of interest is the critical value $D^{s}_\text{0K}$
of $D$ for which an asymmetric $K$-hot-spot solution branch
bifurcates off of the symmetric branch. To this end, we first calculate
from (\ref{2:h0}) that
\bsub \label{2:vl}
\begin{equation}
   v(l) = \frac{(\gamma-\alpha)}{2\alpha^2 \sqrt{D_0}} 
    B \left({l/q}\right) \,,
 \qquad q\equiv \left( \frac{D_0 \pi^2 \alpha^2}{(\gamma-\alpha)^3}
  \right)^{1/4} \,,\label{2:vl_1}
\end{equation}
where the function $B(z)$ on $0<z<\infty$ is defined by
\begin{equation}
   B(z) \equiv z^2 + {1/z^2} \,. \label{2:vl_2}
\end{equation}
\esub The function $B(z)>0$ in (\ref{2:vl_2}) has a unique global
minimum point at $z=z_c=1$, and it satisfies $B^{'}(z)<0$ on $[0,z_c)$
and $B^{\p}(z)>0$ on $(z_c,\infty)$. Therefore, given any $z \in
(0,z_c)$, there exists a unique point $\zt \in (z_c, \infty)$ such
that $B(z)=B(\zt)$. This shows that given any $l$, with ${l/q}
<z_c=1$, there exists a unique $\lt$, with ${\lt/q} \equiv \zt
>z_c=1$, such that $v(l)=v(\lt)$.

We refer to solutions of length $l$ and $\lt$ as A-type and B-type
hot-spots.  Now consider the interval $x \in [a,b]$ with length
$S\equiv b-a$.  To construct a $K$-hot-spot steady-state solution to
(\ref{2:1d}) on this interval with $K_1 \geq 0$ hot-spots of type A and
$K_2 = K-K_1 \geq 0$ hot-spots of type B, arranged in any order across
the interval, we must solve the coupled system $2K_1 l + 2 K_2 \lt=S$
and $B({l/q})=B({\lt/q})$ for $l\neq \lt$.  Such solution exists only
if $l/q<z_c$ and ${\lt/q} > z_c$ with $z_c=1.$ The bifurcation point
corresponds to the minimum point where $l=\lt=q$. With
$D={D_0/\eps^2}$, this yields that
\begin{equation}
   l= \left( \frac{D_0 \pi^2 \alpha^2}{(\gamma-\alpha)^3}
  \right)^{1/4}
   = \eps^{1/2} \left( \frac{D \pi^2 \alpha^2}{(\gamma-\alpha)^3}
  \right)^{1/4} \,.
  \label{scruffyguy}
\end{equation}
At this value of the parameters, a steady-state $K$-hot-spot asymmetric
solution branch bifurcates off of the symmetric $K$-hot-spot branch. This
critical value of $D_0$ determines the small eigenvalue stability
threshold in the linearization of the symmetric $K$-hot-spot steady-state
solution. For a symmetric configuration of $K$ hot-spots on an
interval of length $S$ we have $2Kl=S$ so that the critical value
$D_0=D_{0K}^S$, as defined by (\ref{scruffyguy}), can be written as
\begin{equation}
   D_{0K}^S = \frac{(\gamma-\alpha)^3}{\pi^2 \alpha^2}
   \left(\frac{S}{2K}\right)^4\,.
  \label{2:d0asy}
\end{equation}
A more detailed construction of the asymmetric solution branches
parallels that done in \cite{mjww_2} for the Schnakenburg model and is
left to the reader.

\section{The NLEP Stability of Steady-State 1-D Hot-Spot Patterns}\label{3:stab}

We now study the stability of the $K$-hot-spot steady-state solution to
(\ref{2:1d}) that was constructed in \S \ref{1:eq}.  The analysis for
the ``large'' ${\mathcal O}(1)$ eigenvalues in the spectrum of the
linearization is done in several distinct steps.

Firstly, we let $A_e$, $v_e$ denote the one-hot-spot quasi-steady-state
solution to (\ref{2:1d}) on the basic interval $|x|\leq l$, which was
given in Principal Result 2.1. Upon introducing the perturbation
\begin{equation}
   A = A_e + \phi e^{\lambda t} \,, \qquad 
   v = v_e + \psi e^{\lambda t} \,, \label{3:pert}
\end{equation}
we obtain from the linearization of (\ref{2:1d}) that
\bsub \label{3:eig}
\begin{gather}
\eps^{2} \phi_{xx}-\phi+3\eps^2 v_e A_e^{2}\phi+\eps^3 A_e^{3}\psi = 
  \lambda \phi \,,  \label{3:eig_1} \\
 D_0\left(  \eps A_e^2 \psi_x + 2 A_e v_{ex} \phi \right)_x -
 3\eps^2 A_e^2 v_e \phi - \eps^3 \psi A_e^3 = \tau \lam \eps^2 \left(
 \eps A_e^2 \psi + 2 A_e v_e \phi\right) \,. 
  \label{3:eig_2}
\end{gather}
We consider (\ref{3:eig_1}) and (\ref{3:eig_2}) on $|x|\leq l$
subject to the Floquet-type boundary conditions
\begin{equation}
  \phi(l)=z\phi(-l) \,, \qquad
  \phi^{\p}(l)=z\phi^{\p}(-l) \,, \qquad
  \psi(l)=z\psi(-l) \,, \qquad 
  \psi^{\p}(l)=z\psi^{\p}(-l) \,, \label{3:eig_3}
\end{equation}
\esub 
where $z$ is a complex parameter. 

For simplicity, in this section we will set $\tau=0$ in
(\ref{3:eig_2}). The analysis of the possibility of Hopf bifurcations
induced by taking $\tau\neq 0$ is studied in \S \ref{4:hopf}.

After formulating the NLEP associated
with solving (\ref{3:eig}) for arbitrary $z$, we then must determine
$z$ so that we have the required NLEP problem for a $K$-hot-spot pattern
on $[-l,(2K-1)l]$ with periodic boundary conditions. This is done by
translating $\phi$ and $\psi$ from the interval $[-l,l]$ to the
extended interval $[-l,(2K-1)l]$ in such a way that the extended
$\phi$ and $\psi$ have continuous derivatives at
$x=l,3l,\ldots,(2K-3)l$.  It follows that
$\phi\left[(2K-1)l\right]=z^K\phi(-l)$, and hence to obtain periodic
boundary conditions on an interval of length $2Kl$ we require that
$z^K=1$, so that
\begin{equation}
    z_j = e^{2\pi i j/K} \,, \qquad j=0,\ldots,K-1 \,. \label{3:zj}
\end{equation}
By using these values of $z_j$ in the NLEP problem associated with
(\ref{3:eig}), we obtain the stability threshold of a $K$-hot-spot
solution on a domain of length $2Kl$ subject to periodic boundary
conditions. The last step in the analysis is then to extract the
stability thresholds for the corresponding Neumann problem from the
thresholds for the periodic problem, and to choose $l$ appropriately
so that the Neumann problem is posed on $[-1,1]$. This is done below.
This Floquet-based approach to determine the NLEP problem of a
$K$-hot-spot steady-state solution for the Neumann problem has been used
previously for reaction-diffusion systems exhibiting mesa patterns
\cite{mckay}, for the Gierer-Meinhardt model \cite{vpd}, and for a
cross-diffusion system \cite{kw}.

We now implement the details of this calculation. The asymptotic analysis 
for $\eps\to 0$ of (\ref{3:eig}) proceeds as follows. 
In the inner region with $y=\eps^{-1}x$, we use $A_e={\mathcal O}(\eps^{-1})$
and $v_{ex}\ll 1$, to obtain from (\ref{3:eig_2}) that to leading order
$\left[ w^2 \psi_y\right]_y=0$ in the inner region, where $w$ is the
homoclinic satisfying (\ref{2:iw0}). To prevent exponential growth for
$\psi$ as $|y|\to\infty$, we must take $\psi=\psi_0$ where $\psi_0$ is a
constant to be determined. Then, for (\ref{3:eig_1}) we look for a localized
inner eigenfunction in the form
\begin{equation*}
   \phi \sim \Phi(y) \,, \qquad y=\eps^{-1} x \,.
\end{equation*}
Upon using the leading-order approximation $A_{e}\sim \eps^{-1} v_{0}^{-1/2}w$
in (\ref{3:eig_1}), we obtain to leading order that $\Phi(y)$ satisfies 
\begin{equation}
  L_0 \Phi + \frac{1}{v_{0}^{3/2}} w^3 \psi_0 = \lam \Phi
 \,, \qquad -\infty<y<\infty\,; 
 \qquad  L_0 \Phi \equiv \Phi^{\p\p} - \Phi + 3 w^2 \Phi \,, \label{3:Phi}
\end{equation}
with $\Phi\to 0$ as $|y|\to \infty$. Here $v_0$ is given in (\ref{2:v0v1}).

In the outer region, away from the hot-spot centered at $x=0$, we have
$A_e\sim \alpha$ and $v={\mathcal O}(1)$, so that (\ref{3:eig_1})
yields
\begin{equation}
  \phi = \frac{\eps^3 \alpha^3}{\lambda + 1} \psi \,. \label{3:phi_out}
\end{equation}
Then, from (\ref{3:eig_2}), together with $A_e\sim \alpha$, we obtain the 
outer approximation 
$D_0\left[\eps \alp^2 \psi_x + {\mathcal O}(\eps^3)\right]_x=
{\mathcal O}(\eps^3)$, which yields the leading-order outer problem
\begin{equation}
  \psi_{xx}=0 \,, \qquad 0<|x|\leq l \,, \label{3:psi_out}
\end{equation}
subject to the Floquet-type boundary conditions (\ref{3:eig_3}). The
matching condition for the inner and outer representations of $\psi$
is that $\lim_{x\to 0}\psi(x)=\psi_0$, where $\psi_0$ is the unknown
constant required in the spectral problem (\ref{3:Phi}). However, the
problem for $\psi$ is not yet complete, as it must be supplemented by
appropriate jump conditions for $\psi_x$ across $x=0$.

We now proceed to derive this jump condition. We first define an intermediate
scale $\eta$ satisfying $\eps\ll \eta\ll 1$, and we integrate (\ref{3:eig_2})
over $|x|\leq \eta$ to get
\begin{equation}
  D_0 \left( \eps A_e^2 \psi_x + 2A_e v_{ex} \phi \right)\vert_{-\eta}^{\eta}
 = \int_{-\eta}^{\eta} \left( 3\eps^2 A_e^2 v_e \phi + 
 \eps^3 \psi A_e^3 \right) \, dx \,. \label{3:j1}
\end{equation}
We use the limiting behavior as $x\to 0^{\pm}$ of the outer expansion to 
calculate the terms on the left hand-side of (\ref{3:j1}). From
$A_e\sim \alp$, (\ref{2:h0}) to calculate $v_{ex}(0^{\pm})$,
and (\ref{3:phi_out}) to calculate $\phi(0^{\pm})$, we obtain that
\bsub \label{3:j2}
\begin{gather}
  D_0 \left( \eps A_e^2 \psi_x\right)\vert_{-\eta}^{\eta} \sim 
   D_0 \eps \alp^2 \left( \psi_x(0^{+})- \psi_x(0^{-})\right) = 
  \eps D_0 \alp^2 \left[ \psi_x\right]_0 \,, \label{3:j2_1} \\
  D_0 \left( 2A_e v_{ex} \phi \right)\vert_{-\eta}^{\eta} \sim
  2D_0 \alp \left[ \phi(0^{+}) v_{ex}(0^{+}) -\phi(0^{-}) v_{ex}(0^{-})
  \right] = 4D_0\alp \phi(0^{+}) v_{ex}(0^{+}) \sim 
   \frac{4\eps^3 \alp^2}{\lam +1} \psi(0) (\gamma-\alpha) l \,.
  \label{3:j2_2} 
\end{gather}
\esub 
Here we have defined $\left[\psi_x\right]_0\equiv
\psi_x(0^{+})-\psi_{x}(0^{-})$.

Next, since $\eta\gg {\mathcal O}(\eps)$, we can estimate the integrals on
the right-hand side of (\ref{3:j1}) by their contributions from the inner
approximation $A_e\sim \eps^{-1} v_{0}^{-1/2}w(y)$, $\psi\sim \psi_0$,
$\phi\sim \Phi(y)$, and $v_e\sim v_0$.  In this way, we calculate
\begin{equation}
 \int_{-\eta}^{\eta} \left( 3\eps^2 A_e^2 v_e \phi + 
 \eps^3 \psi A_e^3 \right) \, dx \sim 3\eps \int_{-\infty}^{\infty} w^2 \Phi\, dy
 + \frac{\eps \psi_0}{v_0^{3/2}} \int_{-\infty}^{\infty} w^3 \, dy \,.
   \label{3:j3}
\end{equation}
Upon substituting (\ref{3:j2}) and (\ref{3:j3}) into (\ref{3:j1}), we 
obtain the following jump condition for $\psi_x$ across $x=0$:
\begin{equation}
  D_0 \alp^2 \left[\psi_x\right]_0 = \psi_0 \left( 
  v_0^{-3/2} \int_{-\infty}^{\infty} w^3\, dy - \frac{4\eps^2\alp^2}{\lam +1}
 (\gamma-\alpha) l \right) + 3 \int_{-\infty}^{\infty} w^2 \Phi \, dy \,.
  \label{3:j4}
\end{equation}

For the range $\lambda > -1$, we can neglect the negligible ${\mathcal
  O}(\eps^2)$ term in the jump condition (\ref{3:j4}). In this way,
the problem for the outer eigenfunction $\psi(x)$ is to solve
\bsub \label{3:psi_e}
\begin{equation}
  \psi_{xx}=0 \,, \qquad 0<|x| \leq l \,; \qquad
    \psi(l)=z\psi(-l) \,, \quad
  \psi^{\p}(l)=z\psi^{\p}(-l) \,, \label{3:nojump} \\
\end{equation} 
subject to the continuity condition $\psi(0^{+})=\psi(0^{-})=\psi_0$ and
the following jump condition across $x=0$:
\begin{equation}
   a_0 \left[\psi_{x}\right]_0 + a_1 \psi(0) = a_2 \,; \qquad
  a_0 \equiv D_0 \alp^2 \,, \qquad a_1 = - v_0^{-3/2} \int_{-\infty}^{\infty}
 w^3 \, dy \,, \qquad a_2 = 3\int_{-\infty}^{\infty} w^2 \Phi \, dy \,.
  \label{3:jump_s}
\end{equation}
\esub
Upon calculating $\psi(0)=\psi_0$ from this problem, the NLEP is then
obtained from (\ref{3:Phi}).

Upon solving (\ref{3:psi_e}) for $\psi(x)$, and evaluating the result
at $x=0$ we get
\begin{equation}
  v_0^{-3/2} \psi_0 = -3 \left( \frac{ \int_{-\infty}^{\infty} w^2 \Phi \, dy}
  {\int_{-\infty}^{\infty} w^3 \, dy} \right)
  \left[ 1 - \left( \frac{D_0 \alp^2 v_0^{3/2}}{2 l 
  \int_{-\infty}^{\infty} w^3 \, dy} \right) \frac{(z-1)^2}{z} \right]^{-1}\,.
 \label{3:psi_form}
\end{equation}
Next, we use (\ref{2:v0v1}) for $v_0$ and 
$\int_{-\infty}^{\infty} w^3\, dy=\sqrt{2}\pi$ to simplify $v_0^{-3/2}\psi_0$.
In addition, we use (\ref{3:zj}) to calculate
\begin{equation}
  \frac{(z-1)^2}{z} = -2 + 2 \mbox{Re}(z) = -2\left[ 1 - \cos\left(
  {2 \pi j/K} \right) \right] \,, \quad j=0,\ldots,K-1 \,.\label{3:z}
\end{equation}
Upon substituting these results into (\ref{3:Phi}), we obtain the
following NLEP for a $K$-hot-spot steady-state on a domain of length $2Kl$
subject to periodic boundary conditions:
\bsub \label{3:nlep_p}
\begin{gather}
  L_0 \Phi - \chi_j w^3 \frac{ \int_{-\infty}^{\infty} w^2 \Phi \, dy}
  {\int_{-\infty}^{\infty} w^3 \, dy} = \lam \Phi \,, \qquad -\infty< y<\infty
 \,; \qquad \Phi\to 0 \,, \quad |y|\to \infty \,, \label{3:nlep_p_1}\\
  \chi_j \equiv 3 \left[ 1 + \frac{D_0\alp^2\pi^2} {4l^4(\gamma-\alp)^3}
   \left( 1 - \cos\left({2 \pi j/K} \right)\right) \right]^{-1} \,, \quad j=0,
  \ldots,K-1 \,. \label{n:nlep_p_2}
\end{gather}
\esub

The final step in the analysis is extract the NLEP for the Neumann
problem from the NLEP (\ref{3:nlep_p}) for the periodic problem. More
specifically, the stability thresholds for a $K$-hot-spot pattern with
Neumann boundary conditions can be obtained from the corresponding
thresholds for a $2K$-hot-spot pattern with periodic boundary conditions
on a domain of twice the length.  To see this, suppose that $\phi$ is
a Neumann eigenfunction on the interval $[0,a]$. Extend it by an even
reflection about the origin to the interval $[-a,a]$. Such an
extension then satisfies periodic boundary conditions on $[-a,a]$.
Alternatively, if $\phi(x)$ is an eigenfunction with periodic boundary
conditions at the edge of the interval $[-a,a]$, then define $\hat{\phi}(x)=
\phi(x)+\phi(-x)$. Then, $\hat{\phi}$ is a eigenfunction for
the Neumann boundary problem on $[0,a]$.

Therefore, to obtain the NLEP problem governing the stability of an
steady-state $K$-hot-spot pattern on an interval of length
$S$ subject to Neumann boundary conditions, we simply replace 
$\cos({2\pi j/K})$ with $\cos({\pi j/K})$ in (\ref{3:nlep_p}) and
then set $l=S/(2K)$ in the NLEP of (\ref{3:nlep_p}). In this way, we obtain
the following main result:

\noindent {\bf \underline{Principal Result 3.1}}: {\em Consider a
  $K$-hot-spot solution to (\ref{2:1d}) on an interval of length $S$
  subject to Neumann boundary conditions. For $\eps\to 0$, and
  $\tau={\mathcal O}(1)$, the stability of this solution with respect
  to the ``large'' eigenvalues $\lam=O(1)$ of the linearization is
  determined by the spectrum of the NLEP} \bsub \label{3:nlep_n}
\begin{gather}
  L_0 \Phi - \chi_j w^3 \frac{ \int_{-\infty}^{\infty} w^2 \Phi \, dy}
  {\int_{-\infty}^{\infty} w^3 \, dy} = \lam \Phi \,, \qquad -\infty<
  y<\infty \,; \qquad \Phi\to 0 \,, \quad |y|\to \infty
  \,, \label{3:nlep_n_1}\\ \chi_j = 3 \left[ 1 + \frac{D_0\alp^2\pi^2
      K^4}{4(\gamma-\alp)^3} \left( \frac{2}{S}\right)^4 
   \left( 1 - \cos\left({\pi j/K} \right)\right)
      \right]^{-1} \,, \quad j=0,\ldots,K-1
    \,, \label{n:nlep_n_2}
\end{gather}
\esub
{\em where $w(y)$ is the homoclinic solution satisfying $w^{\p\p}-w+w^3=0$.}

The stability threshold for $D_0$ is characterized by the largest
possible value of $D_0$ for which the point spectrum of
(\ref{3:nlep_n}) satisfies $\mbox{Re}(\lam)<0$ for each
$j=0,\ldots,K-1$. In contrast to the typical NLEP problem associated
with spike patterns in the Gierer-Meinhardt, Gray-Scott, and
Schnakeneburg reaction-diffusion models studied in \cite{dgk_1},
\cite{dgk_2}, \cite{iww}, \cite{kww_1}, \cite{mjww_1}, and
\cite{mjww_2}, \cite{ww_4}, the point spectrum for the
non-self-adjoint problem (\ref{3:nlep_n}) is real, and can be
determined analytically. This fact, as we now show, relies critically
on the identity $L_0 w^2 = 3w^2$ from (\ref{2:l0_iden}).

\noindent {\bf \underline{Lemma 3.2}}: {\em Consider the NLEP problem}
\begin{equation}
  L_0 \Phi - c w^3 \int_{-\infty}^{\infty} w^2 \Phi \, dy = 
    \lam \Phi \,, \qquad -\infty<y<\infty\,; \qquad \Phi\to 0 \,, 
  \quad |y|\to \infty \,, \label{3:lemma}
\end{equation}
{\em for an arbitrary constant $c$ corresponding to eigenfunctions for
  which $\int_{-\infty}^{\infty} w^2 \Phi \, dy\neq 0$. Consider the
  range $\mbox{Re}(\lam)>-1$. Then, on this range there is only one
  element in the point spectrum, and it is given explicitly by}
\begin{equation}
   \lambda = 3 - c \int_{-\infty}^{\infty} w^5 \, dy \,. \label{3:ps}
\end{equation}

To prove this we consider only the region $\mbox{Re}(\lam)>-1$, where
we can guarantee that $|\Phi|\to 0$ exponentially as $|y|\to
\infty$. The continuous spectrum for (\ref{3:lemma}) is $\lam<-1$,
with $\lam$ real.  To establish (\ref{3:ps}) we use Green's identity
on $w^2$ and $\Phi$, which is written as $\int_{-\infty}^{\infty}
\left( w^2 L_0 \Phi - \Phi L_0 w^2\right)\, dy=0$.  Since
$L_0\Phi=cw^3\int_{-\infty}^{\infty} w^2\Phi\, dy + \lam\Phi$ and
$L_0w^2=3w^2$, this identity reduces to
\begin{equation*}
   \int_{-\infty}^{\infty} w^2 \Phi \, dy \left( \lam -3 + c\int_{-\infty}^{\infty}
  w^5 \, dy \right) = 0 \,,
\end{equation*}
from which the result (\ref{3:ps}) follows. We remark that for the
corresponding local eigenvalue problem $L_0 \Phi=\nu\Phi$, it was
proved in Proposition 5.6 of \cite{dgk_0} that the point spectrum
consists only of $\nu_0=3$ and the translation mode $\nu_1=0$ (with
odd eigenfunction), and that there are no other point spectra in
$-1<\nu<0$. When $c=0$, we observe that (\ref{3:ps}) agrees with
$\nu_0$.  As a further remark, the result (\ref{3:ps}), when
extrapolated into the region $\lam<-1$, suggests that there is a
critical value of $c$ for which the discrete eigenvalue bifurcates out of
the continuous spectrum into the region $\lam>-1$ on the real axis.

By applying Lemma 3.2 to the NLEP (\ref{3:nlep_n}) we conclude
that $\mbox{Re}(\lam)<0$ if and only if
\begin{equation}
   \chi_j < 3 \left(  \frac{  \int_{-\infty}^{\infty} w^3 \, dy}
    {\int_{-\infty}^{\infty} w^5 \, dy} \right) =2 \,, \qquad j=0,\ldots,K-1
  \,. \label{3:chi_s}
\end{equation}
In obtaining the last equality in (\ref{3:chi_s}) we calculated the
integrals using $w=\sqrt{2}\sech{y}$. Since $\chi_0=3<\chi_1<\chi_2<\ldots
\chi_{K-1}$, a one-hot-spot solution is stable for all $D_0$, while the
instability threshold for a multi hot-spot pattern is set by
$\chi_{K-1}$. In this way, we obtain the following main stability result.

\noindent {\bf \underline{Principal Result 3.2}}: {\em Consider a
  $K$-hot-spot solution to (\ref{2:1d}) on an interval of length
 $S$ with $K>1$
  subject to Neumann boundary conditions. For $\tau=0$, and
 in the limit $\eps\to 0$, this solution is stable on an ${\mathcal O}(1)$
 time-scale provided that $D_0<D^{L}_\text{0K}$, where}
\begin{equation}
     D^{L}_\text{0K} \equiv \frac{2(\gamma-\alpha)^3 \left({S/2}\right)^4}
  {K^4 \alpha^2 \pi^2
  \left[ 1 + \cos\left({\pi/K}\right)\right]} \,. \label{3:d0large}
\end{equation}
{\em In terms of the original diffusivity $D$, given by $D=\eps^{-2}D_0$,
the stability threshold is $D^{L}_\text{K}=\eps^{-2}D^{L}_\text{0K}$ when
$K>1$. Alternatively, a one-hot-spot solution is stable for all $D_0>0$, 
provided that $D_0$ is independent of $\eps$.}

Although we have not calculated the stability threshold for the small
eigenvalues for which $\lam\to 0$ as $\eps\to 0$ in the spectrum of
the linearization (\ref{3:eig}), we conjecture that this stability
threshold is the same critical value $D^{S}_\text{0K}$ of $D_0$, given in
(\ref{2:d0asy}), for which an asymmetric $K$-hot-spot steady-state branch
bifurcates off of the symmetric $K$-hot-spot branch. This simple approach to
calculate the small eigenvalue stability threshold, which avoids the
lengthy matrix manipulations of \cite{iww}, has been validated for the
Gierer-Meinhardt, Gray-Scott, and Schnakenburg reaction-diffusion
models in \cite{mjww_3}, \cite{mjww_2}, and \cite{kww_3}.
Since $D^{S}_\text{0K}<D^{L}_\text{0K}$ for $K\geq 2$, we conclude that a
symmetric $K$-hot-spot steady-state solution is stable with respect to both 
the large and the small eigenvalues only when $D<D^{S}_\text{0K}$.

We make two remarks. Firstly, for the case of a single hot-spot where
$K=1$ we expect that the stability threshold for $D_0$ will be
exponentially large in $1/\varepsilon$, and similar to that derived in
\cite{kshad} for the Gierer-Meinhardt model in the near-shadow
limit. Secondly, we remark that the possibility of stabilizing
multiple hot-spots for (\ref{2:1d_1}) is in direct contrast to the
result obtained in the analysis of \cite{sww} of spike solutions for
a Keller-Segel-type chemotaxis model with a logarithmic sensitivity
function for the drift term. For this chemotaxis
problem of \cite{sww}, only a one-spike solution can be stable.

\subsection{Numerical Results} \label{3:num}

\begin{figure}[tb]
\begin{center}%
\[
\includegraphics[width=0.95\textwidth]{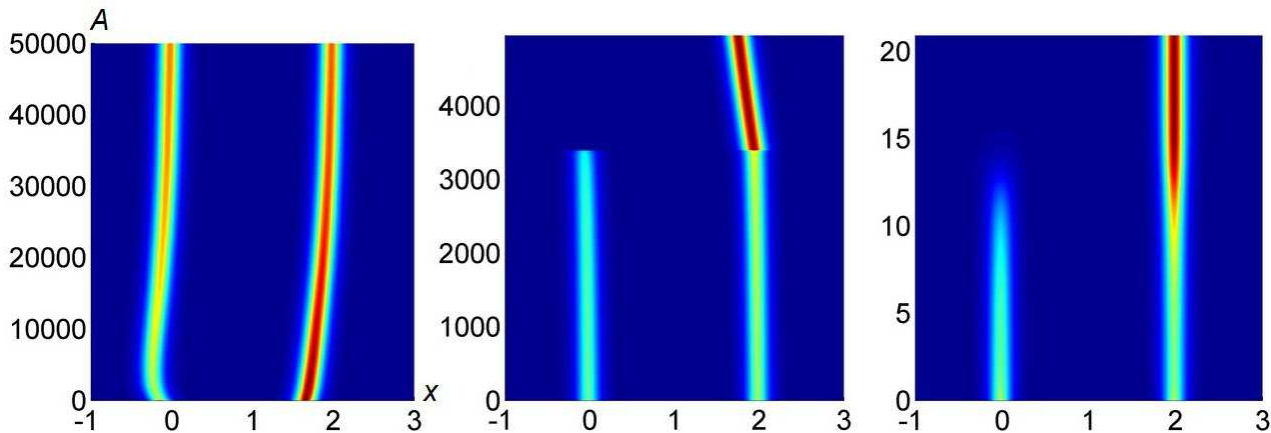}
\]
\end{center}
\caption{Instabilities of a two-hot-spot steady-state solution induced
  by increasing $D.$ Left: two hot-spots are stable with $D=15.$
  Middle: two hot-spots exhibit a slow-time instability when $D=30.$
  Right: there is a fast-time instability when $D=50$.  The parameter values
  are fixed at $\eps=0.07$ $\alpha=1$, and $\gamma=2$, on the interval
  $x\in [-1,3]$. The initial condition for the full numerical solution
  of (\ref{2:1d_2}) consists of two hot-spots that are perturbed
  slightly from the steady-state locations. }%
\label{fig:crime-dyn}%
\end{figure}

We now compare our stability predictions with results from full numerical
solutions of (\ref{2:1d_2}). As derived above, under Neumann
boundary conditions the thresholds on $D$ for the stability 
of a symmetric $K$-hot-spot pattern on a domain of length $2KL$ are
\begin{equation}
D^{S}_\text{K}  \sim \frac{L^{4}}{\eps^2}\frac{\left(\gamma-\alpha\right)^{3}}
 {\alpha^{2}\pi^{2}} \,, \qquad D^{L}_{0K} = D^{S}_{0K}
 \left(  \frac{2}{1+\cos\left( {\pi/K} \right) }\right) \,.
\label{3:mstab_new}
\end{equation}
To numerically validate these thresholds, we choose $\eps=0.07$,
$\alpha=1$, $\gamma=2$, $L=1$ and $K=2$, so that we have an interval
of length $S=4$.  For these parameters, our predicted stability
thresholds are $D^{S}_\text{K}\approx 20.67$ and
$D^{L}_\text{K}\approx 41.33$, and our initial condition is a
two-hot-spot solution with hot-spot locations slightly perturbed from their
steady-state values. For our full numerical solutions of
(\ref{2:1d_2}) we choose either $D=15$, $D=30$, or $D=50$. Our
stability theory predicts the following; the two hot-spots are stable
when $D=15$; the two hot-spots are unstable with respect to only the
small eigenvalues when $D=30$; the two hot-spots are unstable with
respect to both the small and large eigenvalues when $D=50$. The full
numerical results shown in Fig.~\ref{fig:crime-dyn} confirm this
prediction from the asymptotic theory.

\section{Hopf Bifurcation of $K$-Hot-Spot Steady-State Solutions}\label{4:hopf}

In this section we study the spectrum of (\ref{3:eig}) for
$\tau>0$. This is done by first deriving an NLEP similar to
(\ref{3:nlep_n}).  Since the analysis leading to the new NLEP is
very similar to that in \S \ref{3:stab}, we will only
outline it here briefly.

For $\tau\ll {\mathcal O}(\eps^2)$, we get $\psi\approx \psi_0$ in the
inner region $x={\mathcal O}(\eps)$, and hence (\ref{3:Phi}) for the
inner approximation $\Phi(y)$ for $\phi$ remains valid. For $\tau\ll
{\mathcal O}(\eps^2)$, we get to leading-order that $\psi_{xx}=0$ in
the outer region $0<|x|\leq l$ and so (\ref{3:psi_out}) still
holds. However, for $\tau\neq 0$, the jump conditions
(\ref{3:j1})--(\ref{3:j3}) must be modified. In place of
(\ref{3:j1}), we get
\begin{equation}
  D_0 \left( \eps A_e^2 \psi_x + 2A_e v_{ex} \phi \right)\vert_{-\eta}^{\eta}
 = \int_{-\eta}^{\eta} \left( 3\eps^2 A_e^2 v_e \phi + 
 \eps^3 \psi A_e^3 \right) \, dx + \eps^2 \tau\lam \int_{-\eta}^{\eta}
 \left(\eps A_e^2 \psi + 2 A_e v_e \phi\right) \, dx \,. \label{4:j1}
\end{equation}
The left hand-side of (\ref{4:j1}) was estimated in (\ref{3:j2}),
while the first two terms on the right-hand side of (\ref{4:j1}) were
estimated in (\ref{3:j3}). We then use $A_e\sim \eps^{-1}
v_{0}^{-1/2}w(y)$, $\psi\sim \psi_0$, $\phi\sim \Phi(y)$, and $v_e\sim
v_0$, to estimate the last term on the right hand-side of
(\ref{4:j1}) as
\begin{equation}
  \eps^2 \tau\lam \int_{-\eta}^{\eta} \left( \eps A_e^2 \psi + 
  2 A_e v_e \phi\right) \, dx \sim \eps^2\tau \lam \left[
  \frac{\psi_0}{v_0} \int_{-\infty}^{\infty} w^2\, dy + 2\sqrt{v_0} 
 \int_{-\infty}^{\infty} w \Phi \, dy \right] \,. \label{4:j2}
\end{equation}
Upon substituting (\ref{3:j2}), (\ref{3:j3}), and (\ref{4:j2}), into
(\ref{4:j1}), we obtain that 
\begin{equation}
  D_0\eps\alp^2 \left[\psi_x\right]_0 + {\mathcal O}(\eps^3) = \eps
 \left[ 3\int_{-\infty}^{\infty} w^2 \Phi \, dy + \frac{\psi_0}{V_0^{3/2}}
  \int_{-\infty}^{\infty} w^3\, dy \right] + \eps^2 \tau\lam
 \left[\frac{\psi_0}{v_0} \int_{-\infty}^{\infty} w^2\, dy + 2\sqrt{v_0} 
 \int_{-\infty}^{\infty} w \Phi \, dy \right]\,, \label{4:j3}
\end{equation}
which suggests the distinguished limit $\tau=
{\mathcal O}(\eps^{-1})$. Upon defining $\tau_0={\mathcal O}(1)$
by
\begin{equation}
  \tau = \eps^{-1}\tau_0 \,, \label{4:tau_old}
\end{equation}
(\ref{4:j3}) yields the jump condition (\ref{3:jump_s}) for $\psi$
across $x=0$, where $a_0$, $a_1$, and $a_2$ in (\ref{3:jump_s}) are to be
replaced by
\begin{equation}
  a_0 = D\alp^2 \,, \qquad a_1=-v_0^{-3/2} \int_{-\infty}^{\infty} w^3\, dy
 - \frac{\tau_0 \lam}{v_0} \int_{-\infty}^{\infty} w^2 \, dy 
 \,, \qquad a_2 = 3\int_{-\infty}^{\infty} w^2 \Phi \, dy + 2\sqrt{v_0}
 \tau_0\lam \int_{-\infty}^{\infty} w \Phi\, dy \,. \label{4:new_a}
\end{equation}
With this modification of the coefficients in (\ref{3:jump_s}), the outer
problem for $\psi$ is still (\ref{3:psi_e}).

This problem is readily solved for $\psi(x)$, and we obtain that
$\psi_0=\psi(0)$ is given by
\begin{equation}
  v_0^{-3/2} \psi_0 = -\left[ 3 \left(\frac{ \int_{-\infty}^{\infty} w^2 \Phi \, dy}
  {\int_{-\infty}^{\infty} w^3 \, dy}\right)  + 2\tau_0\lam\sqrt{v_0} 
\left(\frac{\int_{-\infty}^{\infty} w\Phi \, dy}{\int_{-\infty}^{\infty} w^3 \, dy}
 \right) \right]\left[1 - 
  \frac{D_0 \alp^2 \pi^2 (z-1)^2}{8l^4 (\gamma-\alp)^3
 z} + \frac{2\tau_0 \lam}{l(\gamma-\alpha)} \right]^{-1} \,.\label{4:psi_0}
\end{equation}

Finally, upon substituting (\ref{2:v0v1}) and (\ref{3:z}) into
(\ref{4:psi_0}), the NLEP problem for the Floquet problem on $[-l,l]$
follows from (\ref{3:Phi}). As shown in \S \ref{3:stab}, this problem
allows us to readily determine the corresponding NLEP for the Neumann
boundary condition problem on an interval of length $S$. The result is 
summarized as follows:

\noindent {\bf \underline{Principal Result 4.1}}: {\em Let
  $\tau={\mathcal O}(\eps^{-1})$ as $\eps\to 0$ and consider a steady-state
  $k$-hot-spot solution on an interval of length $S$ with Neumann boundary 
  conditions. Define $\tau_c={\mathcal O}(1)$ by
  $\tau=\eps^{-1}{S(\gamma-\alpha)\tau_c/(4K)}$. Then, the stability of a
  symmetric $K$-hot-spot steady-state solution is determined by the NLEP}
\bsub \label{4:tnlep}
\begin{equation}
  L_0 \Phi - 3\chi_j w^3 \left( \frac{\int_{-\infty}^{\infty} w^2
    \Phi\, dy} {\int_{-\infty}^{\infty} w^3 \, dy} \right) -
  \frac{\chi_{1j}}{2} w^3 \int_{-\infty}^{\infty} w \Phi\, dy
  =\lambda \Phi\,, \qquad -\infty<y<\infty \,, \label{4:tnlep_1}
\end{equation}
{\em with $\Phi\to 0$ as $|y|\to\infty$. Here, we have defined
$\chi_j$, $\chi_{1j}$, and $\beta_j$ by}
\begin{equation}
  \chi_j \equiv \frac{1}{\left[\beta_j + \tau_c \lam\right]} \,, \qquad
  \chi_{1j} \equiv (\tau_c\lam) \chi_j \,, \qquad 
  \beta_j \equiv 1 + \frac{D_0\alp^2\pi^2 K^4}{4(\gamma-\alp)^3} 
  \left( \frac{2}{S}\right)^4
  \left( 1 - \cos\left({\pi j/K} \right)\right) \,, \quad j=0,\ldots,K-1
 \,.\label{4:tnlep_2}
\end{equation}
\esub

This NLEP, with two separate nonlocal terms, is significantly different
in form from the NLEP's derived for the Gierer-Meinhardt and Gray-Scott
models studied in \cite{dgk_0}, \cite{dgk_1}, \cite{dgk_2}, \cite{mjww_1},
 and \cite{kww_1}.

\noindent \textbf{\underline{Principal Result 4.2}}: \emph{There is no 
value of $\tau_c>0$ for which the NLEP of (\ref{4:tnlep}) has a Hopf
bifurcation.}

\begin{figure}[t]
\begin{center}
\includegraphics[width=0.6\textwidth]{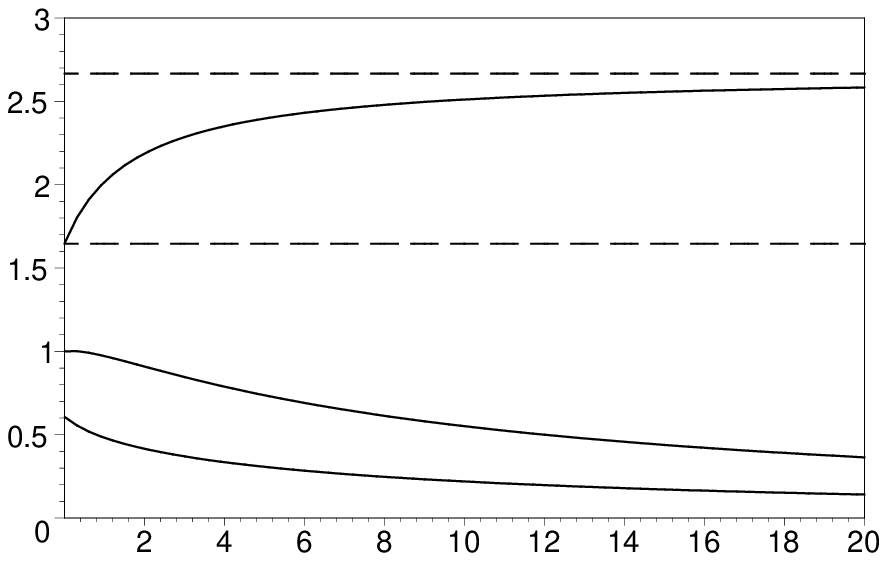}
\end{center}
\setlength{\unitlength}{1\textwidth} \begin{picture}(0,0)(0,0)
\put(-0.005, 0.293){\vector(-1, 1){0.036}}
\put(0, 0.29){$\rho(\mu)
=
\frac{\int w\left[ \left( L_{0}^{2}+\mu \right) ^{-1}\right] L_{0}w\,dy
}{\int w\left( L_{0}^{2}+\mu \right) ^{-1}w\,dy}
$}
\put(-0.13, 0.18){
$\int w\left[ \left( L_{0}^{2}+\mu \right) ^{-1}\right] L_{0}w\,dy
$}
\put(-0.13, 0.18){\vector(-1, -1){0.03}}
\put(0.06, 0.13){
$\int w\left( L_{0}^{2}+\mu \right) ^{-1}w\,dy$
}
\put(0, 0){$\mu$}
\put(0.06, 0.13){\vector(-1, -2){0.025}}

\end{picture}
\caption{A plot of the numerical result for $\rho(\mu)$, as obtained 
from (\ref{wildnight}). Note that $\rho(\mu)$ is monotone increasing.}
\label{fig:rho}
\end{figure}

We note that there is a key step in the derivation of Principal Result
4.2 which relies on a numerical computation, see below. A completely
computer-free derivation of this result is still an open problem.

{\noindent \textbf{\underline{Derivation of Principal Result 4.2:} }} We
use the notation $\int h\,dy\equiv \int_{\infty }^{\infty }h\,dy$. Upon
using Green's identity $\int w^{2}L\Phi \,dy=\int \Phi Lw^{2}\,dy$ and $%
L_{0}w^{2}=3w^{2}$, together with (\ref{4:tnlep_1}), we obtain 
\begin{equation*}
\left( 3-3\chi _{j}\frac{\int w^{5}\,dy}{\int w^{3}\,dy}\right) \int
w^{2}\Phi \,dy-\frac{\chi _{1j}}{2}\left( \int w^{5}\,dy\right) \left( \int
w\Phi \,dy\right) ={\lambda }\int w^{2}\Phi \,dy\,.
\end{equation*}%
Upon solving for $\int w^{2}\Phi \,dy$ in terms of $\int w\Phi \,dy$, and
then substituting into (\ref{4:tnlep_1}), we get 
\begin{subequations}
\label{4:tnlep_n}
\begin{equation}
L_{0}\Phi -f({\lambda })w^{3}\int w\Phi \, dy={\lambda }\Phi \,,\qquad 
 f({\lambda })=\left[ \frac{2}{\chi _{1j}}-\frac{6\chi _{j}}{\chi _{1j}
 (3-{\lambda })}\left( \frac{\int w^{5}\, dy}{\int w^{3}\, dy}\right) 
 \right] ^{-1}\,.
\end{equation}
We then simplify $f(\lambda)$ by using (\ref{4:tnlep_2}), together with 
$\int w^{5}=\left( {3/2} \right) \int w^{3}$, to obtain
\begin{equation}
\frac{1}{f({\lambda })}=\frac{2\beta _{j}}{\tau _{c}{\lambda }}+2-\frac{9}{%
\tau _{c}{\lambda }(3-{\lambda })}\,.  \label{4:fdef}
\end{equation}
\end{subequations}
Next, we observe that the eigenvalues ${\lambda }$ of the NLEP problem
(\ref{4:tnlep_n}) are the roots of the transcendental equation
\begin{equation}
\frac{1}{f({\lambda })}=\int w\left( L_{0}-{\lambda }\right)
^{-1}w^{3}\,dy\,.  \label{4:znlep}
\end{equation}%
Upon recalling that $L_{0}w=2w^{3}$, we calculate
\begin{equation*}
\int w\left( L_{0}-{\lambda }\right) ^{-1}w^{3}\,dy=\frac{1}{2}\int w\left(
L_{0}-{\lambda }\right) ^{-1}\left[ (L_{0}-{\lambda })w+{\lambda }w\right]
\,dy=\frac{1}{2}\int w^{2}\,dy+\frac{{\lambda }}{2}\int w\left( L_{0}-{%
\lambda }\right) ^{-1}w\,dy\,.
\end{equation*}%
Substituting this result together with (\ref{4:fdef}) and $\int w^{2}=4$
into (\ref{4:znlep}), we obtain that ${\lambda }$ is a root of 
\begin{equation}
\frac{{\lambda }}{2}\int w\left( L_{0}-{\lambda }\right) ^{-1}w\,dy=\frac{%
2\beta _{j}}{\tau _{c}{\lambda }}-\frac{9}{\tau _{c}{\lambda }(3-{\lambda })}%
\,.  \label{4:snlep}
\end{equation}

To determine whether a Hopf bifurcation is possible we set ${\lambda }%
=i\lambda _{I}$ in (\ref{4:snlep}) and replace $\left( L_{0}-{\lambda }%
\right) ^{-1}$ by $\left( L_{0}-{\lambda }\right) ^{-1}=\left(
L_{0}^{2}+\lambda _{I}^{2}\right) ^{-1}\left( L_{0}+i{\lambda }_{I}\right) $%
. Then, upon comparing the real and imaginary parts in the resulting
expression, we obtain that $\mu \equiv {\lambda }_{I}^{2}$ and any 
Hopf bifurcation threshold $\tau _{c}$ must be the roots of the coupled system
\begin{equation}
\int w\left[ \left( L_{0}^{2}+\mu \right) ^{-1}\right] L_{0}w\,dy=-\frac{%
4\beta _{j}}{\tau _{c}\mu }+\frac{54}{\tau _{c}\mu (9+\mu )}\,,\qquad \int
w\left( L_{0}^{2}+\mu \right) ^{-1}w\,dy=\frac{18}{\tau _{c}\mu (9+\mu )}\,.
\label{4:fnlep}
\end{equation}%
Upon eliminating $\tau_{c}$ from (\ref{4:fnlep}), we obtain a transcendental
equation solely for $\mu=\lambda _{I}^{2}>0,$%
\begin{equation}
\rho (\mu )=3-2\beta_{j}\text{ }-\frac{2\beta_j}{9}\mu \,,
\text{\ \ \ \ \ where \ \ \ \ }\rho (\mu)\equiv \frac{\int w\left[
    \left( L_{0}^{2}+\mu \right) ^{-1} \right] L_{0}w\,dy}{\int
    w\left( L_{0}^{2}+\mu \right) ^{-1}w\,dy} \,.
\label{wildnight}
\end{equation}%
By using the identities $L_{0} w=2 w^3$ and $L_{0}^{-1} w= {\left(w+
  yw^{\prime}\right)/2}$, the limiting behavior for $\rho(\mu)$ is
readily calculated as
\begin{equation*}
\rho \left( \infty \right) =\frac{\int wL_{0}w\,dy}{\int w^{2}\,dy}=\frac{8}{3}
 \,; \qquad \rho (0)=\frac{\int wL_{0}^{-1}w \, dy}{\int 
 \left( L_{0}^{-1}w\right)^{2} \, dy}= \frac{36}{\pi ^{2}+12}\approx 1.6461 \,.
\end{equation*}%
Moreover, \textbf{direct numerical computations} of $\rho(\mu)$ show that
it is an increasing function of $\mu$ (see Fig.~\ref{fig:rho}). On the other
hand, for $\mu>0$ we have $3-2\beta_{j}$ $-\frac{2\beta_j}{9}\mu <1$
since $\beta_{j}\geq 1.$ It follows that (\ref{wildnight})\ cannot
have any solution with $\mu >0$. Consequently, there is no
Hopf bifurcation on the parameter regime $\tau={\mathcal O}(\eps^{-1})$.
$\blacksquare$

Such a non-existence result for Hopf bifurcations for the crime model
when $\tau={\mathcal O}(\eps^{-1})$ is qualitatively very different
than for the Gierer-Meinhardt and Gray-Scott models, analyzed in
\cite{mjww_1} and \cite{kww_1}, where Hopf bifurcations occur in 
wide parameter regimes.

\subsection{A Hopf Bifurcation for the Shadow Limit}\label{4:hopf_g} 

Principal Result 4.2 has shown that there is no Hopf bifurcation for
the regime $\tau ={\mathcal{O}}({\displaystyle\varepsilon
}^{-1})$. However, a Hopf bifurcation can and does appear when $\tau
={\mathcal{O}}(\varepsilon ^{-2}).$ As will be shown below, in such a
regime the amplitude of the hot-spot becomes oscillatory with an
asymptotically large temporal period, due to an eigenvalue that is
dominated, to leading-order in $\eps$, by its pure imaginary part. To
illustrate this phenomenon, in this section we analytically derive the
condition for a Hopf bifurcation of a single boundary spot on a domain
of length one. To further simplify our computations, we will assume
that $D_{0}$ in (\ref{v})\ is taken sufficiently large such that
$v(x,t)=v(t)$ can be approximated by a time-dependent constant. The
limit $D\to\infty$ is called the shadow-limit
(cf.~\cite{ww_shad}). Our main result is the following:

\noindent \textbf{\underline{Principal Result 4.3}}: \emph{Suppose that }$%
D_{0}\gg 1$\emph{\ and consider a half hot-spot of (\ref{2:1d}) located at
the origin on the domain }$[0,1],$ \emph{as constructed in Principal Result
2.1. Define $\tau_{0c}$ by}%
\begin{equation}
\tau _{0c}=\frac{\left(24-\pi ^{2}\right)}{36\pi ^{2}}\frac{(\gamma -\alpha
)^{3}}{\alpha^2
}=0.039769\frac{(\gamma -\alpha )^{3}}{\alpha^2}.  \label{tau0c}
\end{equation}%
\emph{Let }$\tau =\tau _{0}/\varepsilon ^{2}$\emph{. Then, there is a
  Hopf bifurcation at }$\tau _{0}=\tau _{0c}.$\emph{\ That is, the hot
  spot is stable for }$\tau _{0}<\tau _{0c}$\emph{\ and is unstable
  for }$\tau _{0}>\tau _{0c}.$\emph{\ Destabilization takes place via
  a Hopf bifurcation.  More precisely, when }$\tau
_{0}=O(1),$\emph{\ the related stability problem has an eigenvalue
  near the origin with the following asymptotic behavior as $\eps\to
  0$:}
\begin{equation}
\lambda \sim \left\{ \pm \sqrt{\frac{\gamma -\alpha }{\tau _{0}}}\right\}
i\varepsilon ^{1/2}+\left\{ \frac{\alpha^2 \pi ^{2}}{2(\gamma -\alpha )^{2}}-%
\frac{\left(\gamma -\alpha\right)}{\tau _{0}}\left(
 \frac{24-\pi ^{2}}{72}\right) \right\} \varepsilon \,.  \label{berlin}
\end{equation}

\begin{figure}[t]
\includegraphics[width=1.0\textwidth]{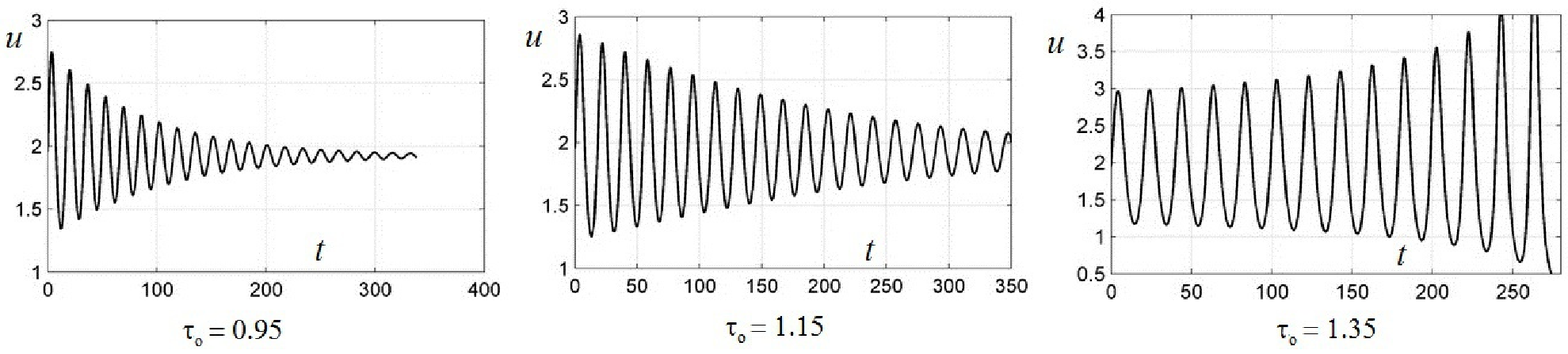} 
\begin{equation*}
(a)~~~~~~~~~~~~~~~~~~~~~~~~~~~~~~~~~~~~~~~~~~~~~~~~
(b)~~~~~~~~~~~~~~~~~~~~~~~~~~~~~~~~~~~~~~~~~~~~~~~~ (c)
\end{equation*}%
\caption{$\max u$ versus $t$ with $\protect\varepsilon
  =0.05,\alpha=1,\protect \gamma =4,\protect\tau =\protect{\tau_{0}
  /\protect\varepsilon^{2}}$ with $ \protect\tau _{0}$ as given in
  the figure. (a) $\protect\tau_{0}<\protect \tau _{c}=1.074;$
  damping is observed. (b) $\protect\tau _{0}>\protect\tau_{c}$ but
  $\protect\tau_{0}$ is very close to $\protect\tau _{c}$; eventual
  fate of the oscillation is unclear. (c) 
  $\protect\tau_{0}>\protect\tau _{c}$; oscillations of increasing
  amplitude are observed.}
\label{fig:osc}
\end{figure}

\textbf{Numerical example.} To illustrate Principal Result 4.3 we take
$\gamma =4,\alpha =1, \varepsilon =0.05$ and $D_{0}=1000.$ Then,
(\ref{tau0c}) yields $\tau _{0c} \approx 1.07378.$ Now take $\tau
_{0}=0.95$ so that (\ref{berlin}) yields the eigenvalue $\lambda
\approx 0.4082i-0.00529.$ We then expect the single hot-spot to be
stable, although it will exhibit long transient oscillations.  From
the eigenvalue, we can estimate the period of the oscillation to be
$P=\frac{2\pi }{0.4082} \approx 15.39.$ This agrees with full
numerical solutions of (\ref{shadow}) as shown in
Fig.~\ref{fig:osc}(a).

Next, we increase $\tau_{0}$ to $1.15$, while keeping the other
parameters the same. In this case, $\tau_{0} > \tau_{0c}$ so that the
hot-spot is unstable in the limit $\varepsilon\to0.$ However,
$\tau_{0} = 1.15$ is very close to the threshold value, and with
$\varepsilon=0.05$, we expect even longer transients with the final
state still unclear at $t=300$. This behavior is shown in
Fig.~\ref{fig:osc}(b).

Finally, as shown in Fig.~\ref{fig:osc}(c), when we increase 
$\tau_{0}$ to $1.35,$ we clearly observe oscillations of an increasing
amplitude.

{\noindent \textbf{\underline{Derivation of Principal Result 4.3:} }}
We begin by re-writing (\ref{2:1d})\ as a shadow system. For
convenience, we also rescale $A$ as $A={u/\eps}$. In terms of this
scaling, $u={\mathcal O}(1)$ in the interior of the hot-spot.
Expanding $v$ in powers of ${\cal D}_0^{-1}$ we then obtain to leading
order that $v(x,t)\sim v(x).$ We then integrate (\ref{v})\ and use the
no-flux boundary conditions to obtain the following shadow-limit
system on $x\in [0,1]$: 
\bsub \label{shadow}
\begin{equation}
u_{t}=\varepsilon ^{2}u_{xx}-u+vu^{3}+\varepsilon \alpha \,, \qquad  \tau
\left( v\int_{0}^{1}u^{2}\, dx\right) _{t}=\mu -\frac{1}{\varepsilon }
v\int_{0}^{1}u^{3}\, dx \,; \qquad u_{x}\left( 0,t\right) =u_{x}(1,t)=0 \,,
\label{shadow_1}
\end{equation}%
where we have defined $\mu$ by
\begin{equation}
\mu \equiv\gamma -\alpha \,.
\end{equation}%
\esub

The shadow problem (\ref{shadow}) is the starting point of our
analysis. The corresponding steady-state system is
\begin{equation}
0=\varepsilon ^{2}u_{xx}-u+vu^{3}+\varepsilon \alpha \,, \qquad
\mu =\frac{1}{\varepsilon }v\int_{0}^{1}u^{3} \, dx \,; \qquad
 u_{x}\left( 0,t\right)=u_{x}(1,t)=0 \,. \label{5:shad_ss}
\end{equation}
As shown below, in order to analyze the Hopf bifurcation it is
necessary to construct the steady-state solution to two orders in
$\varepsilon$. To do so, we let $y={x/\eps}$ and we expand
\begin{equation}
u=u_{0}+\varepsilon u_{1}\ldots \,, \qquad v=v_{0}+\varepsilon v_{1}+\ldots \,.
\end{equation}
By substituting this expansion into (\ref{5:shad_ss}), and equating
powers of $\eps$, we obtain that $u_0=v_{0}^{-1/2} w$, where $w(y)$ is
the positive homoclinic solution of $w_{yy}-w+w^3=0$. In addition,
$u_1$ satisfies
\begin{equation}
 L_{0}u_{1} = -\alpha - w^{3}v_{1}v_{0}^{-3/2} \,, \label{ss:u1}
\end{equation}
where the operator $L_0$ is defined by $L_{0}u \equiv
u_{yy}-u+3w^{2}u$. This operator has several key readily derivable
identities,
\begin{equation*}
L_{0}\left( 1\right) =-1+3w^{2} \,, \qquad L_{0}w=2w^{3} \,, \qquad
L_{0}w^{2}=3w^{2} \,,
\end{equation*}%
which allows us to determine the solution $u_1$ to (\ref{ss:u1}) as
\begin{equation}
u_{1}=\alpha -\alpha w^{2}-\frac{v_1}{2v_0^{3/2}} w \,. \label{ss:u1_sol}
\end{equation}

Next, to determine $v_0$ and $v_1$ we must calculate the integral in
(\ref{5:shad_ss}) for $\eps\ll 1$. This yields that
\begin{equation*}
\mu = \frac{1}{\varepsilon }\int_{0}^{1} vu^{3} \, dx \sim 
\int_{0}^{1/\varepsilon} \left( v_{0}+\varepsilon v_{1}\right) \left(
u_{0}^{3}+3u_{0}^{2}u_{1}\varepsilon \right) \, dy \sim v_{0}^{-1/2}
 \int_{0}^{\infty }w^{3} \, dy+\varepsilon 
\int_{0}^{\infty } \left( 3u_{0}^{2}u_{1}v_{0}+v_{1}u_{0}^{3} \right) \, dy 
 + {\mathcal O}(\eps^2) \,.
\end{equation*}
By equating coefficients of $\eps$, we get that $v_0$ and $v_1$ satisfy
\begin{equation*}
v_{0}^{-1/2}\int_{0}^{\infty }w^{3}dy=\mu  \,, \qquad
\int_{0}^{\infty }3u_{1}w^{2} \, dy  + v_1 v_{0}^{-3/2}\int_{0}^{\infty } w^{3} \,
 dy =0 \,.
\end{equation*}
Upon using the solution (\ref{ss:u1_sol}) for $u_1$, the unknown $v_1$
can be determined in terms of a quadrature as
\begin{equation*}
v_{1}= 6 \alpha v_{0}^{3/2}\frac{\int_{0}^{\infty }\left( w^{2}-w^{4}\right)\,dy }
 {\int_{0}^{\infty }w^{3}\, dy} \,. 
\end{equation*}
The integrals defining $v_0$ and $v_1$ are then calculated explicitly by
using 
\begin{equation*}
w=\sqrt{2}\, \sech{y} \,, \qquad \int_{-\infty }^{\infty }w^{2}\, dy
=4 \,, \qquad \int_{-\infty }^{\infty }w^{3}\, dy =\int_{-\infty
}^{\infty }w \, dy =\sqrt{2}\pi \,, \qquad \int_{-\infty }^{\infty
}w^{4}\, dy =\frac{16}{3}\,,
\end{equation*}
which yields the explicit formulae 
\begin{equation}
v_{0}=\frac{\pi ^{2}}{2}\mu ^{-2}\,, \qquad v_{1}=-2\alpha \pi ^{2}\mu
^{-3} \,. \label{ss:v0v1}
\end{equation}

Next, we study the stability of this solution. For convenience, we
extend the problem to the interval $[-1,1]$ by even reflection.  We
linearize (\ref{shadow}) around the steady-state solution to obtain
the eigenvalue problem
\begin{equation*}
\lambda \phi =\varepsilon ^{2}\phi ^{\prime \prime }-\phi +3u^{2}v\phi
+u^{3}\psi \,; \qquad \qquad \tau \lambda \inte \left( 2\phi
uv+u^{2}\psi \right) \, dy =\frac{-1}{\varepsilon }\inte \left( 3u^{2}v\phi
+u^{3}\psi \right) \, dy \,,
\end{equation*}
where the constant $\psi$ denotes the perturbation in $v$. Upon solving for
$\psi$ we obtain
\begin{equation}
\varepsilon^{2}\phi ^{\prime \prime }-\phi +3u^{2}v\phi
+u^{3}\psi = \lambda \phi \,; \qquad\qquad \psi =
 -\frac{\tau \lambda \varepsilon \inte 2\phi
uv \, dy+\inte 3u^{2}v\phi \, dy }{\tau \lambda \varepsilon \inte u^{2}
\, dy+\inte u^{3} \, dy} \,.
\label{yoga}
\end{equation}%

To motivate the analysis below, we first suppose that $\tau \lambda
\varepsilon \gg 1.$ Then, by using $u\sim {w/\sqrt{v_0}}$ and $v\sim
v_0$ , (\ref{yoga}) reduces to leading order to the NLEP
\begin{equation*}
L_{0}\phi -2w^{3}\frac{\int \phi w}{\int w^{2}} = \lambda \phi \,.
\end{equation*}
Here and below, $\int f$ denotes $\int_{-\infty}^{\infty} f dy.$ This
problem has a zero eigenvalue corresponding to the eigenfunction
$\phi=w$. All other discrete eigenvalues satisfy $\mbox{Re}(\lam)<0$
(cf.~\cite{wei_rev}).  Therefore, the critical eigenvalue will be a
perturbation of the zero eigenvalue.

A posteriori computations shows that the correct anzatz is in fact
\begin{equation*}
\lambda =\varepsilon ^{1/2}\lambda _{0}+\varepsilon \lambda _{1}+\ldots\,;
 \qquad\qquad \tau =\tau_{0}\varepsilon ^{-2} \,.
\end{equation*}%
The analysis below shows that $\lambda _{0}$ is purely imaginary, and
hence determines the frequency of the oscillation, but not its
stability.  Therefore, a two-term expansion in $\lambda $ must be
obtained in order to determine the stability of the oscillations. As
such, we must expand all quantities in the shadow problem up to
${\mathcal O}(\varepsilon )$. The delicate part in the calculation is
to note that
\begin{equation*}
\inte u^{2} =\int u_{0}^{2}  +\varepsilon \int 2u_{0}u_{1}+\varepsilon
^{2}\int_{-1/\varepsilon }^{1/\varepsilon }u_{1}^{2} \,,
\end{equation*}
where the last integral is in fact ${\mathcal O}(\eps)$ as a result of
\begin{equation*}
\varepsilon ^{2}\inte u_{1}^{2}=\varepsilon ^{2}\int_{-1/\varepsilon
}^{1/\varepsilon }(\alpha +\ldots )^{2}=2\alpha^2 \varepsilon +\ldots 
\end{equation*}%
Thus, this term ``jumps'' an order and is comparable in magnitude to
$\eps\int 2 u_0 u_1 $. The remaining part of the analysis is more
straightforward.  We let $\tau=\tau_0\eps^{-2}$ and expand $\psi$ in
 (\ref{yoga}) as
\begin{equation*}
\psi =-\frac{\tau \lambda \varepsilon \int 2\phi uv+\int 3u^{2}v\phi }{\tau
\lambda \varepsilon \int u^{2}+\int u^{3}}=\psi _{0}+\varepsilon ^{1/2}\psi
_{1/2}+\varepsilon \psi_{1} \,,
\end{equation*}%
so that the eigenvalue problem for $\phi$ from (\ref{yoga}) becomes
\begin{align}
(\varepsilon ^{1/2}\lambda _{0}+\varepsilon \lambda _{1})\phi & =L_{0}\phi
+(3u_{0}^{2}v_{1}+6u_{0}u_{1}v_{0})\phi \varepsilon +u_{0}^{3}\left( \psi
_{0}+\varepsilon ^{1/2}\psi _{1/2}+\varepsilon \psi _{1}\right)
+3u_{0}^{2}u_{1}\psi _{0}\varepsilon  \notag \\
& =L_{1}\phi +\varepsilon ^{1/2}L_{2}\phi +\varepsilon L_{3}\phi \,.
  \label{4:yog_1}
\end{align}
After tedious but straightforward computations, the three operators in
(\ref{4:yog_1}) are given by
\bsub \label{4:yog_2}
\begin{align}
L_{1}\phi & \equiv L_{0}\phi -2\frac{\int w\phi }{\int w^{2}}w^{3}\,; 
  \label{4:yog_2a} \\
L_{2}\phi & \equiv \psi _{1/2}u_{0}^{3}=\left( c_{0}\int w\phi +c_{1}\int
w^{2}\phi \right) w^{3}\,; \label{4:yog_2b} \\
L_{3}\phi & \equiv (3u_{0}^{2}v_{1}+6u_{0}u_{1}v_{0})\phi
+3u_{0}^{2}u_{1}\psi _{0}+u_{0}^{3}\psi _{1} \notag \\
& =\left( c_{2}w+c_{3}w^{2}+c_{4}w^{3}\right) \phi +\left(
c_{5}w^{2}+c_{6}w^{3}+c_{7}w^{4}\right) \int w\phi +c_{8}w^{3}\int \phi
+c_{9}w^{3}\int w^{2}\phi \,, \label{4:yog_2c}
\end{align}%
\esub
in terms of the coefficients $c_0,\ldots,c_9$ defined by
\begin{gather}
c_{0} =\frac{\mu }{4\tau _{0}\lambda _{0}}\,; \qquad c_{1}=-\frac{3\mu 
\sqrt{2}}{4\pi \tau _{0}\lambda _{0}}\,; \qquad
c_{2} =\frac{3\pi \alpha \sqrt{2}}{\mu }\,; \qquad  c_{3}=0\,; \qquad
c_{4}=-c_{2} \,; \qquad c_{5} =-{\frac{3\sqrt{2}\pi \alpha }{4\mu }}
\notag \\
 c_{6}=-{\frac{\mu\lambda _{1}}{4\tau _{0}\lambda _{0}^{2}}}-
 {\frac{\mu ^{2}}{8\tau_{0}^{2}\lambda _{0}^{2}}+}\frac{{\pi }^{2}\alpha^2 }
 {8\mu ^{2}}\,; \qquad c_{7}= -c_5 \,; \qquad c_{8} =
 -{\frac{\pi \sqrt{2}\alpha }{4\mu }\,; \qquad
 }c_{9}={\frac{\sqrt{2}\pi \alpha }{4\mu }}+{\frac{\sqrt{2}3\mu 
 \lambda _{1}}{4\pi \tau_{0}\lambda _{0}^{2}}}+
 {\frac{\sqrt{2}3\mu ^{2}}{8\pi \tau _{0}^{2}\lambda
_{0}^{2}}}  \,.  \label{c1-9} 
\end{gather}
Finally, we expand $\phi$ in (\ref{4:yog_1}) as
\begin{equation*}
\phi =w+\varepsilon ^{1/2}\phi _{1}+\varepsilon \phi _{2} \,,
\end{equation*}
and equate powers of $\eps$ in (\ref{4:yog_1}). This yields the following
problems for $\phi_1$ and $\phi_2$:
\begin{align}
\lambda _{0}w& =L_{1}\phi _{1}+L_{2}w  \,, \label{phi1} \\
\lambda _{1}w+\lambda _{0}\phi _{1}& =L_{1}\phi _{2}+L_{2}\phi _{1}+L_{3}w\,.
\label{phi2}
\end{align}%

To determine $\lambda _{0}$ and $\phi _{1}$ we must formulate the
appropriate solvability condition based on the adjoint operator
$L_1^{\ast}$ of $L_1$ defined by
\begin{equation}
L_{1}^{\ast }\phi \equiv L_{0}\phi -2\frac{\int w^{3}\phi }{\int w^{2}}w \,.
  \label{4:adj}
\end{equation}
Since $L_{1}$ admits a zero eigenvalue of multiplicity one, then
so does $L_{1}^{\ast }.$ In fact, $w^{\ast}$ defined by
\begin{equation}
   w^{\ast } \equiv \left( yw_{y}+w\right) /2 \,, \label{4:adj_1}
\end{equation}
is the unique element in the kernel of $L_{1}^{\ast}$, i.e.
$L_{1}^{\ast} w^{\ast}=0$, owing to the following two readily derived
identities:
\begin{equation}
L_{0}w^{\ast }=w \,, \qquad  2\frac{\int w^{3}w^{\ast }}{\int w^{2}}=1 \,.
\label{4:adj_2}
\end{equation}

Next, we impose a solvability condition on (\ref{phi1}) in the usual
way.  We multiply (\ref{phi1}) by $w^{\ast }$ and integrate by parts
to derive that
\begin{equation}
\lambda_{0} =\frac{\int w^{\ast }L_{2}w}{\int w^{\ast }w}=\left( c_{0}\int
w^{2}+c_{1}\int w^{3}\right) \frac{\int w^{\ast }w^{3}}{\int w^{\ast }w} =
 2\left( c_{0}\int w^{2}+c_{1}\int w^{3}\right) \,, \label{4:lam0}
\end{equation}
where we used the integral identity in (\ref{4:adj_2}) together with
$\int w^{\ast }w=\int w^{2}/4.$ From the formulae for the coefficients
$c_0$ and $c_1$ in (\ref{c1-9}), (\ref{4:lam0}) determines $\lam_0$ as
\begin{equation}
\lambda_{0}=\pm i\sqrt{\frac{\mu }{\tau _{0}}} \,. \label{4:l0}
\end{equation}
Since $\lambda_0$ is purely imaginary, the next order term
$\lambda _{1}$ needs to be computed to determine stability.

The problem (\ref{phi1}) for $\phi_1$ can be written by using 
(\ref{4:yog_2a}) and (\ref{4:yog_2b}) as
\begin{equation*}
L_{0}\phi _{1}=\lambda _{0}w-\left( c_{0}\int w^{2}+c_{1}\int w^{3}+2\frac{
\int w\phi _{1}}{\int w^{2}}\right) w^{3}\,.
\end{equation*}
Since $L_0 w^{\ast}=w$ and $L_0 w= 2w^3$, we can write $\phi_1$ as
\begin{equation}
\phi_{1}=\lambda _{0}w^{\ast }-\frac{d w}{2}\, \qquad \text{where} \qquad
d \equiv c_{0}\int w^{2}+c_{1}\int w^{3}+2\frac{\int w\phi _{1}}{\int w^{2}}\,.
 \label{4:phi_1s}
\end{equation}%
Upon substituting $\phi_1$ into the definition of $d$, we can then solve
for $d$ by using (\ref{4:lam0}) for $\lam_0$ to get
\begin{equation*}
2d  =c_{0}\int w^{2}+c_{1}\int w^{3}+2\lambda _{0}\frac{\int ww^{\ast }}{
\int w^{2}} = c_{0}\int w^{2}+c_{1}\int w^{3}+2\left( c_{0}\int w^{2}+c_{1}\int
w^{3}\right) \frac{\int w^{\ast }w^{3}}{\int w^{2}}  \,.
\end{equation*}
Finally, by using $\int w^{\ast }w=\int w^{2}/4$, the expression above 
simplifies to
\begin{equation*}
d  =c_{0}\int w^{2}+c_{1}\int w^{3}
\end{equation*}
so that $\phi_1$ is given explicitly from (\ref{4:phi_1s}) as
\begin{equation}
\phi _{1}=\left( c_{0}\int w^{2}+c_{1}\int w^{3}\right) \left( 2w^{\ast } -
\frac{w}{2}\right) \,. \label{4:phi1_f}
\end{equation}%

With $\phi_1$ explicitly known, we impose the solvability condition on
the problem (\ref{phi2}) for $\phi_2$ to determine $\lam_2$ as
\begin{equation*}
\lambda _{1}=\frac{\int w^{\ast }\left( L_{2}-\lambda _{0}\right) \phi
_{1}+\int w^{\ast }L_{3}w}{\int w^{\ast }w} \,.
\end{equation*}
Upon using (\ref{4:phi1_f}) for $\phi_1$, $L_3 w$ from
(\ref{4:yog_2c}), and $(L_2-\lam_0)$ from (\ref{4:yog_2b}), the integrals
above are evaluated as
\begin{align*}
\lambda _{1}& =\left( -{\frac{16\pi ^{2}}{9}}-{\frac{16}{3}}\right) 
{c}_{{0}}^{2}+\left( -{\frac{4}{3}}\sqrt{2}\pi -
{\frac{8}{9}}\sqrt{2}\pi ^{3}\right) 
{c}_{{1}}{c}_{{0}}-{\frac{2\pi ^{4}}{9}}c_{1}^{2} \\
& +\frac{\sqrt{2}\pi }{3}c_{2}+\frac{3\sqrt{2}\pi }{5}c_{4}+\frac{4\sqrt{2}
\pi }{3}c_{5}+8{c}_{{6}}+\frac{12\sqrt{2}\pi }{5}c_{7}+2\sqrt{2}{c}_{{8}}\pi
+2\sqrt{2}\pi {c}_{{9}} \,.
\end{align*}
Finally, upon substituting $c_{0},\ldots,c_9$ from (\ref{c1-9}) into this
expression, we determine $\lam_1$ explicitly as
\begin{equation}
 \lambda _{1}=\frac{
\alpha^2 \pi ^{2}}{2\mu ^{2}}-\frac{\mu }{\tau _{0}}\left( \frac{24-\pi ^{2}}
 {72}\right)   \label{final-lam}
\end{equation}
The two-term expansion for $\lam$ given in (\ref{berlin}) follows from
(\ref{4:l0}) and (\ref{final-lam}).  The Hopf bifurcation threshold is
obtained by setting $\lambda_{1}=0.$ This occurs precisely as 
$\tau_{0}$ is increased past $\tau _{0c}$, where $\tau_c$ is given by
(\ref{tau0c}). $\blacksquare $

\setcounter{equation}{0}
\setcounter{section}{4}
\section{Hot-Spot Patterns in 2-D: Equilibria and Stability}\label{5:multi}

In this section we construct a $K$-spot quasi-steady-state solution to
(\ref{1:main}) in an arbitrary 2-D domain with spots centered at
$x_1,\ldots, x_K$. To leading-order in $\sigma={-1/\log\eps}$, we then
derive a threshold condition on the diffusivity $D$ for the stability
of the $K$-spot quasi-steady-state solution to instabilities that
develop on an ${\mathcal O}(1)$ time-scale.

As in the analysis of hot-spot patterns in one spatial dimension, we
set $V={P/A^2}$ (see (\ref{2:v_intro})) into (\ref{1:main}) to obtain
\bsub\label{5:2D}
\begin{align}
A_t & =\eps^{2}\Delta A - A+ V A^{3}+ \alpha \,, \qquad x\in \Omega \,;
  \qquad \partial_n A=0 \,, \qquad x\in \partial\Omega \,, \label{5:A}\\
\tau\left(A^2 V\right)_t & =D\nabla \cdot \left(  A^{2} \nabla V\right)-
   VA^{3}+ \gamma -\alpha \,, \qquad x\in \Omega \,;
  \qquad \partial_n V=0 \,, \qquad x\in \partial\Omega \,.
 \label{5:V}
\end{align}
\esub

We first motivate the $\eps$-dependent re-scalings of $V$ and $A$ that
are needed for the 2-D case. We suppose that $D\gg 1$, so that $V$ is
approximately constant. By integrating the steady-state equation of
(\ref{5:V}) over $\Omega$ we get $V={c/\int_\Omega A^3\, dx}$, where
$c$ is some ${\mathcal O}(1)$ constant. Therefore, if $A={\mathcal
  O}(\eps^{-p})$ in the inner hot-spot region of area ${\mathcal
  O}(\eps^2)$, we obtain $\int_\Omega A^3 \, dx = {\mathcal
  O}(\eps^{2-3p})$, so that $V={\mathcal O}(\eps^{3p-2})$. In
addition, from the steady-state of (\ref{5:A}), we must have in the
inner region that $A^3 V \sim A$, so that $-3p+(3p-2)=-p$. This yields
$p=2$. Therefore, for $D\gg 1$, $V={\mathcal O}(\eps^4)$ globally on
$\Omega$, while $A={\mathcal O}(\eps^{-2})$ in the inner region near a
hot-spot.  Finally, in the outer region we must have $A\sim \alp
={\mathcal O}(1)$, so that from (\ref{5:V}), we conclude that $D
\nabla \cdot \left(A^2 \nabla V\right) \sim \alpha -\gamma ={\mathcal
  O}(1)$. Since $V={\mathcal O}(\eps^4)$, this balance requires that
$D={\mathcal O}(\eps^{-4})$.  Finally, in the core of a hot-spot
we conclude that the density $P$ of criminals, given by $P=VA^2$, is
${\mathcal O}(1)$ as $\eps\to 0$.

Although this simple scaling analysis correctly
identifies the algebraic factors in $\eps$, there are more subtle
logarithmic terms of the form $\sigma\equiv {-1/\log\eps}$ that are
needed in the construction of the quasi-steady-state hot spot solution.

The scaling analysis above motivates the introduction of new variables
$v$, $u$, and ${\cal D}$ defined by
\begin{equation}
  V = \eps^4 v \,, \qquad A = \eps^{-2} u \,,
  \qquad D = {{\cal D}/\eps^4}  \,. \label{5:new}
\end{equation}
In terms of (\ref{5:new}), (\ref{5:2D}) transforms exactly to
\bsub\label{5:2d}
\begin{align}
 u_t & =\eps^{2}\Delta u - u + v u^3 + \alpha\eps^2  \,, \qquad x\in\Omega \,; 
 \qquad \partial_n u = 0 \,, \qquad x\in\partial\Omega \,, \label{5:2d_1}\\
\tau \left(u^2 v\right)_t & =\frac{\cal D}{\eps^4} 
 \nabla \cdot \left(u^2\nabla v \right) -\eps^{-2} vu^{3}+ \gamma -\alpha 
 \,, \qquad x\in\Omega\,;
 \qquad \partial_n v =0 \,. \qquad x\in\partial\Omega \,. \label{5:2d_2}
\end{align}
\esub 
Owing to the non-uniformity in the behavior as $|y|\to\infty$ of the
solution to this core problem near a spot (see below), the
construction of a quasi-steady-state $K$-spot solution for (\ref{5:2d})
is more intricate than that for the Gierer-Meinhardt, Schnakenburg, or
Gray-Scott problems analyzed in \cite{ww_1}--\cite{ww_5}, \cite{kmjw},
\cite{kww_2}, and \cite{cw}. As such, we will only develop a theory
that is accurate to leading order in $\sigma\equiv {-1/\log\eps}$,
similar to that undertaken in \cite{ww_1}--\cite{ww_5}, and
\cite{kmjw}. This is in contrast to the recent approach in
\cite{kww_2} and \cite{cw} that used a hybrid asymptotic-numerical
method to construct quasi-steady-state spot patterns to the
Schnakenburg and Gray-Scott systems, respectively, with an error that
is beyond-all-orders with respect to $\sigma$.

In the outer region, away from the spots centered at $x_1,\ldots,x_K$, we
expand $u$ and $v$ as
\begin{equation}
   u = \alp \eps^2 + o(\eps^2) \,, \qquad v \sim h_0 + \sigma h_1 + \cdots\,,
  \label{5:out}
\end{equation}
where $\sigma={-1/\log\eps}$ and ${\cal D}={ {\cal D}_0/\sigma}$, where
${\cal D}_0={\mathcal O}(1)$. From the steady-state of (\ref{5:2d_2})
we obtain that $h_0$ is constant, and that $h_1$ satisfies
\begin{equation}
  \Delta h_1 = - \frac{(\gamma-\alp)}{\alp^2 {\cal D}_0} \,, \qquad
  x\in \Omega\backslash \lbrace{x_1,\ldots,x_K\rbrace} \,; \qquad
  \partial_n h_1 = 0 \,, \qquad x\in \Omega \,. \label{5:h1}
\end{equation}
As shown below, this problem must be augmented by certain
singularity conditions that are obtained by matching the outer
solution for $v$ to certain inner solutions, one in the neighborhood
of each spot.

In the inner region near the $j$-th spot centered at $x_j$ we introduce
the inner variables $y$, $u_j$, and $v_j$ by
\begin{equation}
   y=\eps^{-1}(x-x_j) \,, \qquad v_j(y)=v(x_j+\eps y) \,, \qquad
    u_j(y)=u(x_j+\eps y) \,. \label{5:inn_var}
\end{equation}
In terms of these inner variables, and with ${\cal D}=\sigma^{-1} {\cal D}_0$, 
the steady-state of (\ref{5:2d}) transforms exactly on $y\in \R^2$ to
\bsub\label{5:inn}
\begin{gather}
 \Delta_y u_j - u_j + v_j u_j^3 + \alpha\eps^2  = 0 \,, \label{5:inn_1}\\
 \nabla_y \cdot \left(u_j^2 \nabla_y v_j\right) -
  \frac{\eps^4\sigma}  { {\cal D}_0 } v_j u_j^3 = -
  \frac{\eps^6 \sigma} { {\cal D}_0} (\gamma-\alpha) \,,
  \label{5:inn_2}
\end{gather}
\esub
where $\sigma\equiv {-1/\log\eps}$. We will construct a radially symmetric
solution $u_j=u_j(\rho)$, $v_j=v_j(\rho)$ to this problem, where
$\rho=|y|$.

The complication in analyzing (\ref{5:inn}) is that $u_j={\mathcal
  O}(1)$ for $|y|={\mathcal O}(1)$, whereas $u_j={\mathcal O}(\eps^2)$
for $|y|\gg 1$. Therefore, the ``diffusivity" $u_j^2$ in the operator
for $v$ in (\ref{5:inn_2}) ranges from ${\mathcal O}(1)$ when
$|y|={\mathcal O}(1)$ to ${\mathcal O}(\eps^4)$ when $|y|\gg 1$. For
this reason, we cannot simply neglect the second term on the left-hand
side of (\ref{5:inn_2}) for all $|y|$. However, as we show below, we
can neglect the ${\mathcal O}(\eps^6)$ term on the right-hand side of
(\ref{5:inn_2}).

For $|y|={\mathcal O}(1)$, we expand $u_j$ and $v_j$ as
\begin{equation}
    u_j = u_{j0} + \eps^2 u_{j1} + \cdots \,, \qquad
    v_j = v_{j0} + \eps^2 v_{j1} + \cdots \,. \label{5:ujvj}
\end{equation}
Upon substituting this expansion into (\ref{5:inn}), we obtain that
$v_{j0}$ and $v_{j1}$ are constants, and that $u_{j0}$ and $u_{j1}$
are radially symmetric solutions of
\begin{equation*}
   \Delta_\rho u_{j0} - u_{j0} + v_{j0} u_{j0}^3 = 0 \,, \qquad
   \Delta_\rho u_{j1} - u_{j1} + 3u_{j0}^2 v_{j0} u_{j1} = -\alpha -
   u_{j0}^3 v_{j1}\,,
\end{equation*}
on $\rho\geq 0$ with $u_{j0}\to 0$ and $u_{j1}\to \alpha$ as $\rho\to
\infty$. Here $\Delta_{\rho} g \equiv g^{\prime\prime} +\rho^{-1}
g^{\prime}$ for $g=g(\rho)$. In terms of the unknown constants $v_{j0}$ and
$v_{j1}$, the solutions for $u_{j0}$ and $u_{j1}$ are
\begin{equation}
    u_{j0} = v_{j0}^{-1/2} w \,, \qquad u_{j1} = \alpha - 
 \frac{v_{j1}}{2 v_{j0}^{3/2}} w - 3\alpha w_1 \,, \label{5:u0u1}
\end{equation}
where $w=w(\rho)$ and $w_{1}=w_{1}(\rho)$ are the unique radially symmetric
solutions of
\begin{equation}
   \Delta_{\rho} w - w + w^3 = 0 \,, \qquad
  L w_1 \equiv \Delta_{\rho} w_1 - w_1 + 3 w^2 w_1 = w^2 \,, \label{5:ww1}
\end{equation}
with $w(\rho)>0$, $w^{\prime}(0)=0$, and $w\to 0$ as $\rho\to\infty$,
together with $w_{1}^{\prime}(0)=0$ and $w_{1}\to 0$ as
$\rho\to\infty$.  The expression (\ref{5:u0u1}) for $u_{j1}$ shows
that $u_{j1}\to \alpha$ as $\rho\to\infty$, so that from
(\ref{5:ujvj}) $u_{j}\sim \alp \eps^2$ when $|y|\gg 1$.

Next, we calculate the far-field behavior, valid for $|y|\gg 1$, for
the solution $v_j$ to (\ref{5:inn_2}). To do so, we define the ball
${\cal B}_\delta=\lbrace{y \,\, \vert \,\, |y|\leq \delta\rbrace}$,
where $1\ll \delta \ll {\mathcal O}(\eps^{-1})$. Therefore, this ball
is defined in the intermediate matching region between the inner and
outer scales $y$ and $x$, respectively. Upon integrating
(\ref{5:inn_2}) over ${\cal B}_\delta$, and using the divergence
theorem, we obtain that
\begin{equation} \label{5:flux}
   2\pi u_{j}^{2} \delta v_{j}^{\prime} \vert_{\rho=\delta} \sim
  \frac{\eps^4\sigma}{ {\cal D}_0} \int_{{\cal B}_\delta} v_j u_j^3\, dy
 + {\mathcal O}(\delta^2 \sigma \eps^6) \,.
\end{equation}
Since $u_j={\mathcal O}(1)$ only for $|y|={\mathcal O}(1)$ where
$v_j\sim v_{j0}+ o(1)$, the integral on the right hand-side of
(\ref{5:flux}) can be estimated by using $u_{j}\sim v_{j0}^{-1/2}
w$. In contrast, on the left hand-side of (\ref{5:flux}) we use
$u_j\sim\alp\eps^2$ on $\rho=\delta\gg 1$. In this way, we obtain that
(\ref{5:flux}) becomes
\begin{equation}
  2\pi \alp^2 \delta \eps^4 v_{j}^{\prime} \vert_{\rho=\delta} \sim
  \left(\frac{2\pi \eps^4 \sigma }{\sqrt{v_{j0}} \, {\cal D}_0}\right)
  \int_{0}^{\infty} w^3 \, \rho\, d\rho  + 
 {\mathcal O}(\delta^2 \sigma \eps^6) \,.
  \label{5:flux_1}
\end{equation}
Since $\delta\ll {\mathcal O}(\eps^{-1})$, we can neglect the last
term on the right hand-side of (\ref{5:flux_1}), which is equivalent
to neglecting the right-hand side of (\ref{5:inn_2}) for the inner
problem.

From (\ref{5:flux_1}) we obtain that the far-field behavior for $|y|\gg 1$
for the solution to (\ref{5:inn_2}) has the form
\bsub \label{5:ff}
\begin{equation}
  v_{j} \sim v_{j0} + \sigma \left( S_j \log|y| + {\mathcal O}(1) \right)
  \,, \label{5:ff_1}
\end{equation}
where $S_j$ is defined by
\begin{equation}
   S_j \equiv \frac{b}{\alpha^2 {\cal D}_0 \sqrt{v_{j0}}} \,, \qquad
  b\equiv \int_{0}^{\infty} w^3 \rho \, d\rho \,. \label{5:ff_2}
\end{equation}
\esub
Therefore, the appropriate core problem determining the asymptotic
shape of the hot-spot profile is to seek a radially symmetric
solution to
\begin{equation}
 \Delta_y u_j - u_j + v_j u_j^3 + \alpha\eps^2  = 0 \,, \qquad
 \nabla_y \cdot \left(u_j^2 \nabla_y v_j\right) =
  \frac{\eps^4\sigma}  { {\cal D}_0 } v_j u_j^3 \,. \label{5:inn_n}
\end{equation}


The next step in the construction of the multi hot-spot quasi-steady-state
pattern is to match the inner and outer solutions for $v$ in order to
determine $v_{j0}$. We let $y=\eps^{-1}(x-x_j)$ in (\ref{5:ff_1}) to
obtain that the outer solution for $v$ must have the singularity
behavior
\begin{equation}
 v\sim v_{j0} + S_j + \sigma \left[ S_j \log|x-x_j| + {\mathcal O}(1)\right]\,,
 \qquad \mbox{as} \quad x\to x_j \,, \quad j=1,\ldots,K \,. \label{5:sing}
\end{equation}
Upon comparing (\ref{5:sing}) with the outer expansion $v\sim h_0 +
\sigma h_1 + \cdots$ from (\ref{5:out}), we conclude that
\bsub
\begin{equation} \label{5:m1}
  h_0=v_{j0} + S_j \,, \qquad j=1,\ldots,K \,,
\end{equation}
and that $h_1$ satisfies (\ref{5:h1}) subject to the singularity behaviors
\begin{equation}
  h_1 \sim S_j \log|x-x_j| + {\mathcal O}(1) \,, 
  \qquad \mbox{as} \quad x\to x_j \,, \quad j=1,\ldots,K \,. \label{5:m2}
\end{equation}
Upon using the divergence theorem, the problem for $h_1$ has a solution
only when the solvability condition
\begin{equation}
 \sum_{j=1}^{K} S_j = \frac{(\gamma-\alpha)}{2\pi\alp^2 {\cal D}_0} |\Omega|
 \,, \label{5:m3}
\end{equation}
\esub 
is satisfied, where $|\Omega|$ is the area of $\Omega$. When
this condition is satisfied, the solution for $h_1$ can be written as
\begin{equation}
   h_1 = - 2\pi \sum_{i=1}^{K} S_i G(x;x_i) + \bar{h}_1 \,, \label{5:h1sol}
\end{equation}
where $\bar{h}_1$ is a constant to be determined, and where $G(x;x_i)$ is 
the Neumann Green's function satisfying
\bsub \label{5:green}
\begin{gather}
   \Delta G = \frac{1}{|\Omega|} -\delta(x-x_i) \,, \qquad x\in \Omega\,;
 \qquad \partial_n G = 0 \,, \qquad x\in \partial\Omega\,, \\
   \int_{\Omega} G(x;x_i) \, dx = 0 \,; \qquad  G\sim -\frac{1}{2\pi}
  \log|x-x_i| + {\mathcal O}(1) \,, \quad \mbox{as} \quad x\to x_i \,.
\end{gather}
\esub

In summary, the asymptotic matching provides the following algebraic
system for determining $v_{j0}$ for $j=1,\ldots,K$:
\begin{equation}
     h_0 = {\cal F}(v_{j0}) \equiv v_{j0} + \frac{c}{\sqrt{v_{j0}}} \,, 
  \qquad \sum_{j=1}^{K} v_{j0}^{-1/2} = \frac{|\Omega| (\gamma-\alp)}{2\pi b} \,,
  \qquad c \equiv  \frac{b}{\alp^2 {\cal D}_0} \,. \label{5:sys}
\end{equation}
A symmetric $K$-hot-spot quasi-steady-state solution corresponds to a
solution of (\ref{5:sys}) for which $v_{j0}=v_0$ for all $j$. From
(\ref{5:u0u1}), this solution is characterized by the fact that the
hot-spot profile is, to leading-order, the same for each
$j$. The result for such symmetric quasi-equilibria is summarized as
follows:

\begin{figure}[tb]
\begin{center}%
$$
\includegraphics[width=0.4\textwidth]{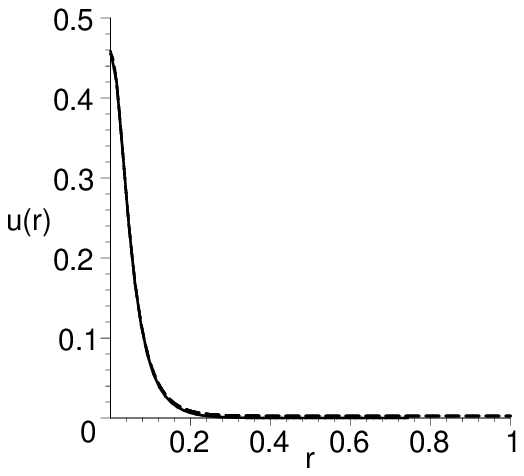}
~~~~~~~~~~~
\includegraphics[width=0.4\textwidth]{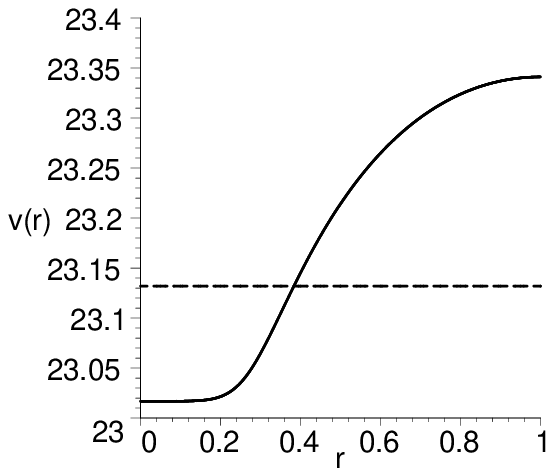}
$$
$$(a)~~~~~~~~~~~~~~~~~~~~~~~~~~~~~~~~~~~~~~~~~~~~~~~~~~~~~~~~~~~~(b)$$
\end{center}
\caption{Steady state hot-spot solution of (\ref{5:2d}) in a unit disk. 
Parameter values are $
\varepsilon=0.05, \alpha=1, \gamma=2, D=500\eps^{-2}, r=|x|<1.$ 
(a) The solid line is the steady state solution $u(r)$ of (\ref{5:2d}) 
computed by solving the associated radially symmetric boundary value problem
numerically. The dashed line is the asymptotic approximation
given by (\ref{5:sum_1}) (b) The solid line is the steady state
solution for $v(r).$ Note the ``flat knee'' region within the spot center.
The dashed line is the leading-order asymptotics $v\sim v_0$ given by 
(\ref{5:sum_1}). } \label{fig:ss2d} \end{figure}

\noindent \textbf{\underline{Principal Result 5.1}}: {\em For $\eps\to
  0$, a symmetric $K$-hot-spot quasi-steady-state solution to
  (\ref{5:2d}) on the parameter regime $D=\eps^{-4}{{\cal
      D}_0/\sigma}$ with $\sigma={-1/\log\eps}$, is characterized as
  follows: In the inner region near the $j$-th hot-spot, where
  $y=\eps^{-1}(x-x_j)$, then}
\begin{equation}
    u \sim v_{0}^{-1/2} w + {\mathcal O}(\eps^2) \,, \qquad
    v \sim v_0 + o(1) \,, \qquad v_0 \equiv \frac{4\pi^2 b^2 K^2}
  {|\Omega|^2 (\gamma-\alp)^2} \,, \label{5:sum_1} 
\end{equation}
{\em where $b=\int_{0}^{\infty} w^3 \rho \, d\rho$, and $w(\rho)$ with
  $\rho=|y|$ is the ground-state solution of (\ref{5:ww1}).
  Alternatively, in the outer region where $|x-x_j|\gg {\mathcal O}(\eps)$
  for $j=1,\ldots,K$, then}
\begin{equation}
   u\sim \alp\eps^2 \,, \qquad v \sim h_0 + \sigma h_1 + \cdots \,.
\end{equation}
{\em Here $h_1$ is given in (\ref{5:h1sol}) in terms of the Neumann Green's
function, and $h_0$ is a constant given by}
\begin{equation}
   h_0 = v_0 + \frac{b}{\alp^2 {\cal D}_0 \sqrt{v_0}} = 
  \frac{4\pi^2 b^2 K^2} {|\Omega|^2 (\gamma-\alp)^2 }  + 
  \frac{|\Omega| (\gamma-\alpha) }{2\pi\alp^2 K {\cal D}_0} \,. \label{5:h0_sum}
\end{equation}
{\em For a symmetric hot-spot pattern, the source strength $S_j$ in
(\ref{5:h1sol}) is the same for each $j$, and is given in terms of
$v_0$ by $S_j = S_0 \equiv {b/\left(\alpha^2 {\cal D}_0 \sqrt{v_{0}}\right)}$
where $b\equiv \int_{0}^{\infty} w^3 \rho \, d\rho$.}

To illustrate Principal Result 5.1, we let $\Omega$ be the unit disk
centered at the origin, and we choose $\alpha=1, \gamma=2,
\varepsilon=0.05$, and $D=500\varepsilon^{-2}.$ We consider a single
hot-spot at the center of the disk.  Upon numerically computing the
ground-state solution $w$ satisfying (\ref{5:ww1}) we obtain that
\begin{equation*}
w(0) \approx 2.2062 \,, \qquad  \int_{\mathbb{R}^{2}} w^{3}\, dy \approx 15.1097
 \,.
\end{equation*}
The asymptotic result (\ref{5:sum_1}) then yields
\[
v_{0}\sim
\frac{\left(15.1097\right)^2}{
\left(\gamma-\alpha\right)^2 \pi^2}
\approx
23.13184;\ \ \ \ \ u(0)\sim w(0)v_{0}^{-1/2}\approx0.4587.
\]
Alternatively, from the full numerical solution of the radially symmetric
steady-state solution of (\ref{5:2d}), we compute that
$v(0) \approx 23.017$  and $u(0) \approx 0.455.$ The
error is about 0.5\% for $v(0)$ and about 0.8\% for $u(0).$
A comparison of asymptotic and numerical results is shown in
Fig.~\ref{fig:ss2d}.

As a remark, the general shape of the function ${\cal F}(v)$ defined
in (\ref{5:sys}) also shows that there can be asymmetric $K$-spot
quasi-equilibria corresponding to spots of two distinct
heights. Similar asymmetric patterns in 2-D have been constructed for
the Gierer-Meihnardt and Gray-Scott systems in \cite{ww_4}) and
(\cite{ww_5}. Let $K_s$ and $K_b$ be non-negative integers denoting
the number of small and large spots, respectively, with
$K=K_s+K_b$. Then, from (\ref{5:sys}), a $K$-spot asymmetric pattern
is constructed by determining two distinct values $v_{0s}$ and
$v_{0b}$ satisfying
\begin{equation}
   {\cal F}(v_{0l})={\cal F}(v_{0r})  \,, \qquad v_{0b}< v_\text{min}<v_{0s} \,,
 \qquad \frac{K_s}{\sqrt{v_{0s}}} + \frac{K_b}{\sqrt{v_{0b}}} =
\frac{|\Omega| (\gamma-\alp) }{2\pi b} \,, \label{5:asym}
\end{equation}
where $v_\text{min}=3\left({c/2}\right)^{3/2}$ and $c$ is defined in
(\ref{5:sys}). The spatial  profile of the hot-spot of large and small
amplitude is given by $u\sim v_{0s}^{-1/2}w$ and $u\sim v_{0b}^{-1/2}
w$, respectively.

We will not investigate the solvability with respect to ${\cal D}_0$
of the algebraic system (\ref{5:asym}) governing asymmetric spot
patterns.  Instead, in the next subsection we will study the
stability properties of the symmetric $K$-hot-spot quasi-steady-state
solution given in Principal Result 5.1.

\subsection{The Stability of Hot-Spot Quasi-Steady-State Patterns}\label{5:stab}

We linearize (\ref{5:2d}) around the quasi-steady-state $K$-hot-spot pattern
to obtain the eigenvalue problem
\bsub \label{5:spec}
\begin{gather}
\eps^{2}\Delta\phi-\phi+3u^{2}v\, \phi+u^{3}\, \psi=\lam \phi \,, \qquad
 x\in \Omega\,; \qquad \partial_n \phi=0 \,, \qquad x\in \partial\Omega \,,
  \label{5:spec_1}\\
 D\nabla\cdot\left(u^{2}\nabla\psi +  2u\phi\nabla v\right)  -\frac
{1}{\eps^{2}}\left(  3u^{2}v\, \phi+u^{3}\, \psi\right) =\tau\lam \left(
 u^2 \psi + 2uv \phi\right) \,, \qquad
 x\in \Omega\,; \qquad \partial_n \psi=0 \,, \qquad x\in \partial\Omega \,.
  \label{5:spec_2}
\end{gather}
\esub 
In our stability analysis we will consider the range of $D$ where
$D=\eps^{-4}{{\cal D}_0/\sigma}$ and $D_0={\mathcal O}(1)$. It is on
this parameter range of $D$ that a stability threshold occurs.  Our main
stability result is as follows:

\noindent \textbf{\underline{Principal Result 5.2}}: {\em Consider the
  $K$-spot quasi-steady-state solution of (\ref{5:2d}) as constructed
  in Principal Result 5.1 for $\eps\ll 1$.  Assume that
  $\tau={\tau_0/\eps^2}$ where $\tau_0={\mathcal O}(1)$. Then, for
  $\eps\to 0$, and to leading order in $\sigma={-1/\log\eps}$, the
  stability on an ${\mathcal O}(1)$ time-scale of a $K$-hot-spot
  quasi-steady-state solution with $K\geq 2$ is determined by the
  spectrum of the two distinct NLEP's}
\bsub \label{5:nlep}
\begin{equation}
   \Delta_y \Phi - \Phi + 3 w^2 \Phi - \kappa_i w^3 \left[
  \frac{\int_{\R^2} w^2 \Phi \, dy}{\int_{\R^2} w^3 \, dy} + 
  \frac{2}{3} \tau_0\lam \beta
  \frac{\int_{\R^2} w \Phi \, dy}{\int_{\R^2} w^2 \, dy} \right] =\lam\Phi \,,
  \qquad y\in \R^2 \,, \label{5:nlep_1}
\end{equation}
{\em with $\Phi\to 0$ as $|y|\to\infty$. Here $\kap_i$ for $i=1,2$ 
are defined by}
\begin{equation}
   \kap_1 \equiv 3\left( 1 + \tau_0\lam \beta \right)^{-1} \,, \qquad
  \kap_2 \equiv 3 \left( 1 + \tau_0\lam \beta + \frac{{\cal D}_0}{2{\cal D}_{0c}}
  \right)^{-1} \,,
  \label{5:nlep_2}
\end{equation}
{\em in terms of the constants $\beta$ and ${\cal D}_{0c}$ defined by}
\begin{equation}
  \beta \equiv \frac{K \int_{\R^2} w^2\, dy}{|\Omega|(\gamma-\alp)} \,,
  \qquad {\cal D}_{0c} \equiv \frac{|\Omega|^3 \left(\gamma-\alp\right)^3}
  {4\pi \alp^2 K^3 \left(\int_{\R^2} w^3 \, dy\right)^2}
  \,. \label{5:nlep_3}
\end{equation}
\esub {\em For a one-hot-spot solution, for which $K=1$, there is only
  a single NLEP with $\kappa_i$ in (\ref{5:nlep_1}) replaced by
  $\kappa_1$.  For $\tau_0\ll 1$, which is equivalent to $\tau\ll
  {\mathcal O}(\eps^{-2})$, we conclude that a $K$-spot pattern with
  $K\geq 2$ is stable when ${\cal D}_0<{\cal D}_{0c}$ and
  is unstable when ${\cal D}_0>{\cal D}_{0c}$. In terms of the
  unscaled $D$, this yields the stability threshold}
\begin{equation}
D = \left(\frac{\eps^{-4}}{\sigma}\right) {\cal D}_{0c} \,, \quad \mbox{for}
 \quad K\ge 2 \,, \qquad \sigma = {-1/\log\eps} \,.
\end{equation}
{\em For $K=1$ and $\tau_0\ll 1$, a single hot-spot
  is stable for all ${\cal D}_0$ independent of $\eps$.}

To show that ${\cal D}_{0c}$ is the stability threshold of ${\cal D}_0$ for 
$K\geq 2$, we let $\tau_0\to 0$ in (\ref{5:nlep}) to obtain the
limiting NLEP's
\begin{equation}
   \Delta_y \Phi - \Phi + 3 w^2 \Phi - \kappa w^3 
  \frac{\int_{\R^2} w^2 \Phi \, dy}{\int_{\R^2} w^3 \, dy} =\lam\Phi \,,
  \qquad y\in \R^2 \,; \qquad \Phi\to 0 \quad \mbox{as} \quad |y|\to\infty\,,
 \label{5:snlep_1}
\end{equation}
where $\kappa$ can assume either of the two values
\begin{equation}
   \kap_{10} \equiv 3\,, \qquad \kap_{20} \equiv 3 \left( 1 + 
 \frac{{\cal D}_0}{2{\cal D}_{0c}}\right)^{-1} \,.
  \label{5:snlep_2}
\end{equation}
For (\ref{5:snlep_1}), the rigorous result in Theorem 3 of
\cite{zhang} establishes the existence of an eigenvalue with
$\mbox{Re}(\lam)>0$ whenever $\kappa<2$. Thus, since $\kap_{20}<2$
when ${\cal D}_0>{\cal D}_{0c}$, we conclude that a $K$-hot-spot
quasi-equilibria with $K>2$ is unstable when ${\cal D}_0>{\cal
  D}_{0c}$. In addition, the result in Theorem 1 of \cite{zhang} (see
the remark following Theorem 1) proves that all eigenvalues
satisfy $\mbox{Re}(\lam)<0$ when $2<\kappa\leq 6$. Since
$\kap_{10}=3$ and $2<\kap_{20}<3$ for all ${\cal D}_0<{\cal D}_{0c}$,
it follows that the NLEP for a $K$-hot-spot quasi-steady-state solution 
with $K\geq 2$ does not have unstable eigenvalues when 
${\cal D}_0<{\cal D}_{0c}$. For $K=1$, which corresponds to a one-hot-spot 
solution, the only choice for $\kappa$ is $\kappa=\kap_{10}=3$, and so we 
have stability for any ${\cal D}_0>0$ independent of $\eps$.

We make two remarks. Firstly, we anticipate that a single hot-spot
solution where $K=1$ will be stable provided that $D_0$ is not
exponentially large in $1/\varepsilon$. The analysis to determine this
stability threshold in the near-shadow limit should be similar to that
done for the Gierer-Meinhardt model in \cite{kshad}.  Secondly, it is
an open question to analyze (\ref{5:nlep}) for $\tau_0={\mathcal
  O}(1)$ to determine if there are any Hopf bifurcations. The analysis
of this problem in the 2-D context is much more difficult than in 1-D
since the identity $L_{0}w^2 = 3w^2$ for the local operator
$L_0\Phi\equiv \Delta_y \Phi - \Phi + 3w^2 \Phi$ in $\R^1$ no longer
holds in $\R^2$. Moreover, since this identity does not hold in the 2-D
case, we cannot determine $\lambda$ explicitly when $\tau_0\ll 1$ as was
done for the 1-D case in Lemma 3.2.

The stability threshold in Principal Result 5.2 follows once we derive
the NLEP (\ref{5:nlep}). The derivation of this NLEP is done in several
distinct steps.

We first consider the inner region near the $j$-th spot and we
introduce the new variables $y$, $\Phi_j(y)$, and $\Psi_j(y)$ by
\begin{equation*}
 y=\eps^{-1}(x-x_j) \,, \qquad \Phi_j(y)=\phi(x_j+\eps y) \,, \qquad
 \Psi_j(y)=\psi(x_j+\eps y) \,.
\end{equation*}
Then, with $D=\eps^{-4}{{\cal D}_0/\sigma}$ and $\sigma={-1/\log\eps}$,
(\ref{5:spec}) on $y\in \R^2$ becomes
\bsub \label{5:Spec}
\begin{gather}
   \Delta_y \Phi_j - \Phi_j + 3 u_j^2 v_j \, \Phi_j + u_j^3 \, \Psi_j = \lam
  \Phi_j \,, \label{5:Spec_1} \\
 \nabla_y\cdot\left(  u_j^{2}\nabla_y\Psi_j + 2u_j \Phi_j\nabla_y v_j\right)
 -\frac{\eps^4\sigma}{{\cal D}_0} \left(  3u_j^{2}v_j\, \Phi_j+u_j^{3}\,
 \Psi_j\right) =\frac{\tau\lam\eps^6\sigma}{{\cal D}_0} \left(
 u_j^2 \Psi_j + 2u_jv_j \Phi_j\right) \,.  \label{5:Spec_2}
\end{gather}
\esub

Since $u_j={\mathcal O}(1)$ and $\nabla_y v_j = o(1)$ when
$|y|={\mathcal O}(1)$, we obtain from (\ref{5:Spec_2}) that
$\Psi_j=\Psi_{j0}+{\mathcal O}(\eps^2)$ when $|y|={\mathcal O}(1)$,
where $\Psi_{j0}$ is an unknown constant. Then, with $u_j\sim
{w/\sqrt{v_0}}$ and $v_{j}\sim v_0$ for $|y|={\mathcal O}(1)$, we
obtain from (\ref{5:Spec_1}) that $\Phi_j\sim \Phi_{j0}+o(1)$, where
$\Phi_{j0}$ satisfies
\begin{equation}
   \Delta_y \Phi_{j0} - \Phi_{j0} + 3w^2 \Phi_{j0} + v_{0}^{-3/2} w^3 \,
  \Psi_{j0} = \lam \Phi_{j0} \,, \qquad y \in \R^2\,; \qquad
  \Phi_{j0} \to 0  \quad \mbox{as} \quad |y|\to \infty \,,
  \label{5:Phi_j}
\end{equation}
where $w$ is the radially symmetric ground-state solution satisfying
(\ref{5:ww1}).

The next step in the analysis is to determine the constant $\Psi_{j0}$ in
(\ref{5:Phi_j}). This is done by first determining the far-field behavior
as $|y|\to \infty$ of the solution to (\ref{5:Spec_2}). We define the
ball ${\cal B}_\delta=\lbrace{y \,\, \vert \,\, |y|\leq \delta\rbrace}$,
where $1\ll \delta \ll {\mathcal O}(\eps^{-1})$, and we integrate
(\ref{5:Spec_2}) over ${\cal B}_\delta$ to obtain
\begin{equation}
  \left(2\pi u_j^2 \Psi_j^{\prime}\vert_{\rho=\delta} + 4\pi
 u_j \Phi_j v_{j}^{\prime} \vert_{\rho=\delta} \right) \delta = 
   \frac{\eps^4\sigma}{{\cal D}_0}  \int_{{\cal B}_\delta} \left(
 \Psi_j u_j^3 + 3 u_j^2 v_j \Phi_j \right) \, dy 
  + \frac{\tau\lambda \eps^6 \sigma}{{\cal D}_0} 
   \int_{{\cal B}_\delta} \left( \Psi_j  u_j^2 + 2 u_j v_j \Phi_j \right)\, dy
   \,.  \label{5:ball_1}
\end{equation}
We now estimate the terms in (\ref{5:ball_1}) for $\eps\ll 1$.

Since the dominant contribution to the integrals on the right hand-side
of (\ref{5:ball_1}) arises from the region where $|y|={\mathcal O}(1)$,
we can asymptotically estimate these integrals by using
$u_j\sim {w/\sqrt{v_0}}$, $v_j\sim v_0$, and $\Psi_j\sim\Psi_{j0}$. For the
left hand-side of (\ref{5:ball_1}) we use $u_j\sim \alp\eps^2$ on 
$\rho\equiv |y|=\delta\gg 1$ to get
\begin{multline}
 \left(2\pi \eps^4 \alp^2 \Psi_{j}^{\prime} \vert_{\rho=\delta} + 
 4\pi \eps^2 \alp v_{j}^{\prime} \Phi_j \vert_{\rho=\delta} \right) \delta 
  \sim \frac{\eps^4\sigma}{{\cal D}_0} \left[
  \frac{\Psi_{j0}}{v_0^{3/2}} \int_{\R^2} w^3 \, dy + 3 
  \int_{\R^2} w^2 \Phi_{j0} \, dy \right] \\
   + \frac{\tau\lam \eps^6 \sigma}{{\cal D}_0} \left[
  \frac{\Psi_{j0}}{v_0} \int_{\R^2} w^2 \, dy + 2 \sqrt{v}_0 
   \int_{\R^2} w\Phi_{j0} \, dy \right] \,. \label{5:ball_2}
\end{multline}

Next, we estimate the second term on the left hand-side of
(\ref{5:ball_2}).  For $|y|\gg 1$, we obtain from the outer limit of
(\ref{5:Spec_1}) that $-\Phi_j + \eps^3 \alp^6 \Psi_j\sim \lam
\Phi_j$, so that with $\Psi_j\sim \Psi_{j0}+o(1)$, we get
\begin{equation}
   \Phi_{j}\sim \frac{\eps^6 \alp^3}{1+\lam} \Psi_{j0} \,, \qquad
  \mbox{on} \quad \rho=|y|=\delta\gg 1 \,. \label{5:phiout}
\end{equation}
In addition, from (\ref{5:ff_1}), we estimate that
$v_{j}^{\prime}\vert_{\rho=\delta}\sim {\sigma S_0/\delta}$, where
$S_0$ is defined in Principal Result 5.1.  Substituting this estimate
together with (\ref{5:phiout}) into (\ref{5:ball_2}) we obtain
\begin{multline}
  \Psi_{j}^{\prime} \vert_{\rho=\delta}\delta + 
  \frac{2\alp^2 \eps^4 \sigma S_0}{(1+\lam)} \Psi_{j0} \sim
\frac{\sigma}{2\pi {\cal D}_0 \alp^2} \left[
  \frac{\Psi_{j0}}{v_0^{3/2}} \int_{\R^2} w^3 \, dy + 3 
  \int_{\R^2} w^2 \Phi_{j0} \, dy \right] \\
   + \frac{\tau\lam \eps^2 \sigma}{2\pi {\cal D}_0 \alp^2} \left[
  \frac{\Psi_{j0}}{v_0} \int_{\R^2} w^2 \, dy + 2 \sqrt{v}_0 
   \int_{\R^2} w\Phi_{j0} \, dy \right] \,. \label{5:ball_f}
\end{multline}

From (\ref{5:ball_f}), and under the assumption that
$\tau=\eps^{-2}\tau_0$, where $\tau_0={\mathcal O}(1)$, we conclude 
that $\Psi_j$ has the far-field behavior
\begin{equation}
  \Psi_j \sim \Psi_{j0} + \sigma \left( B_j \log|y| + {\mathcal O}(1) \right)
 \,, \qquad \mbox{for} \quad |y|\gg 1 \,, \label{5s:ff}
\end{equation}
where $B_j$ for $j=1,\ldots,K$ is defined by
\begin{equation}
  B_j \sim \frac{1}{2\pi {\cal D}_0 \alp^2} \left[
  \frac{\Psi_{j0}}{v_0^{3/2}} \int_{\R^2} w^3 \, dy + 3 
  \int_{\R^2} w^2 \Phi_{j0} \, dy \right] 
   + \frac{\tau_0\lam }{2\pi {\cal D}_0 \alp^2} \left[
  \frac{\Psi_{j0}}{v_0} \int_{\R^2} w^2 \, dy + 2 \sqrt{v}_0 
   \int_{\R^2} w\Phi_{j0} \, dy \right] \,. \label{5s:bj}
\end{equation}
Upon writing (\ref{5s:ff}) in terms of the outer variable $(x-x_j)=\eps y$,
we obtain that the matching condition for the outer solution $\psi$ is
\begin{equation}  
   \psi \sim B_j + \Psi_{j0} + \sigma \left( B_j \log|x-x_j| + {\mathcal O}(1)
  \right) \,, \qquad \mbox{as} \quad x\to x_j \,, \quad j=1,\ldots,K \,.
  \label{5s:ff_1}
\end{equation}

Next, we consider the outer region for $\psi$ where $|x-x_j|={\mathcal O}(1)$
for $j=1,\ldots,K$. In (\ref{5:spec_2}) we use $u\sim \alp \eps^2$,
$\nabla v = {\mathcal O}(\sigma)$, and $\phi={\mathcal O}(\eps^6)$ from
(\ref{5:phiout}) to estimate that
\begin{equation*}
   \frac{{\cal D}_0}{\sigma \eps^4} \nabla \cdot \left(
 \alp^2 \eps^4 \nabla \psi + {\mathcal O}(\sigma \eps^8)\right) -
  \eps^{-2} \left( \alp^3 \eps^6 \psi + {\mathcal O}(\sigma \eps^{10})\right)
 = \tau\lam \left(\alp^2 \eps^4 \psi +  {\mathcal O}(\sigma \eps^8)\psi
 \right)\,.
\end{equation*}
Hence, when $\tau=\eps^{-2}\tau_0$ we obtain that ${\cal D}_0 \Delta
\psi \sim \tau_0 \lam \eps^2\sigma\psi$. In order to match with (\ref{5s:ff_1})
we must expand the outer solution for $\psi$ as
\begin{equation}
     \psi = \psi_0 + \sigma \psi_1 + \cdots \,. \label{5s:outerpsi}
\end{equation}
We then obtain that $\psi_0$ is a constant, given by
\begin{equation}
 \psi_0 = \Psi_{j0} + B_j \,, \qquad j=1,\ldots,K \,, \label{5s:phi_0}
\end{equation}
and that $\psi_1$ satisfies
\bsub \label{5s:out_1}
\begin{gather}
   \Delta \psi_1 = 0 \,, \qquad x\in \Omega\backslash \lbrace{x_1,\ldots,x_K
 \rbrace} \,; \qquad \partial_n \psi_1=0 \,, \quad x\in \partial\Omega\,,
   \\
  \psi_1 \sim B_j \log|x-x_j| + {\mathcal O}(1) \,, \qquad \mbox{as} \quad
  x\to x_j \,, \quad j=1,\ldots,K \,.
\end{gather}
\esub
The solvability condition for (\ref{5s:out_1}) is that $\sum_{j=1}^{K} B_j=0$.
Upon summing (\ref{5s:phi_0}) from $j=1,\ldots,K$, we obtain that 
\begin{equation}
    B_j = -\Psi_{j0} + \frac{1}{K} \sum_{j=1}^{K} \Psi_{j0} \,, \label{5s:bj_2}
\end{equation}
where $B_j$ is given in (\ref{5s:bj}).

The final step in the derivation of the NLEP is to solve (\ref{5s:bj})
and (\ref{5s:bj_2}) for $\Psi_{j0}$ and substitute the resulting
expression into (\ref{5:Phi_j}). To do so, We introduce the vectors ${\cal
 B}\equiv \left(B_{1},\ldots,B_{K}\right)^T$ and $\hat{\Psi}\equiv
\left(\Psi_{01},\ldots,\Psi_{0K}\right)^T$, and we write the system
(\ref{5s:bj}) and (\ref{5s:bj_2}) in matrix form as
\begin{equation}
   {\cal B} = c_1 \hat{\Psi}_0 - c_2 {\cal F}_0 + c_3 \hat{\Psi}_0 - 
 c_4 {\cal F}_1 \,, \qquad \qquad {\cal B} = - \left(I - {\cal E} \right)
 \hat{\Psi}_0 \,, \label{b:mat_1}
\end{equation}
where the constants $c_i$ for $i=1,\ldots,4$ are defined by
\bsub \label{5s:mat_2}
\begin{equation}
   c_1 \equiv \frac{1}{2\pi {\cal D}_0\alp^2 v_0^{3/2}} \int_{\R^2}
   w^3 \, dy \,, \qquad c_2 \equiv 3 c_1 v_0^{3/2} \,, \qquad c_3 \equiv
   \frac{\tau_0\lam}{2\pi {\cal D}_0 \alp^2 v_0} \int_{\R^2} w^2\, dy
   \,, \qquad c_4 \equiv 2v_0^{3/2} c_3 \,, \label{5s:mat_21}
\end{equation}
and the vectors ${\cal F}_0$ and ${\cal F}_1$ are defined by
\begin{equation}
   {\cal F}_0 \equiv -\frac{\int_{\R^2} w^2 \hat{\Phi}_0 \, dy}{\int_{\R^2} w^3\,
 dy} \,, \qquad
   {\cal F}_1 \equiv -\frac{\int_{\R^2} w \hat{\Phi}_0 \, dy}{\int_{\R^2} w^2\,
 dy} \,. \label{5s:f}
\end{equation}
\esub
Here $\hat{\Phi}_0\equiv \left(\Phi_{01},\ldots,\Phi_{0K}\right)^T$. In
(\ref{b:mat_1}), $I$ is the $K\times K$ identity matrix and the matrix
${\cal E}$ is defined by ${\cal E}=K^{-1}e e^{T}$, where 
$e\equiv (1,\ldots,1)^T$.

By solving (\ref{b:mat_1}) for $\hat{\Psi_0}$, and substituting the
resulting expression into (\ref{5:Phi_j}), we obtain the vector NLEP
\begin{equation}
  \Delta_y \hat{\Phi}_0  - \hat{\Phi}_0 + 3w^2 \hat{\Phi}_0 + 
   w^3 {\cal M} \left( {\cal F}_0 + \frac{c_4}{c_2} {\cal F}_1
  \right) = \lam \hat{\Phi}_0 \,, \label{5:vnlep}
\end{equation}
where the matrix ${\cal M}$ is defined by
\begin{equation}
   {\cal M} \equiv v_{0}^{-3/2} \left[ \frac{(1+c_1+c_3)}{c_2} I -
     \frac{1}{c_2}{\cal E} \right]^{-1} \,. \label{5s:m}
\end{equation}
The matrix ${\cal M}^{-1}$ is a rank-one update of a scalar multiple
of the identity matrix. As such, its spectrum 
${\cal M} \omega = \kap \omega$ can readily be calculated as
\begin{equation*}
   \kap_1 = \frac{c_2}{(c_1+c_3) v_0^{3/2}} \,, \quad
   \omega_1=(1,\ldots,1)^T \,; \qquad \kap_2 = \ldots = \kap_K=
   \frac{c_2}{(1+ c_1 + c_3) v_0^{3/2}} \,, \quad \omega_j^T e=0 \,,
   \quad j=2,\ldots,K\,.
\end{equation*}
Notice that the eigenvectors corresponding to the matrix eigenvalues
$\kap_j$ for $j=2,\ldots,K$ span the $K-1$ dimensional subspace
perpendicular to $e=(1,\ldots,1)^T$. As such, the {\em
  competition instability modes} correspond to $j=2,\ldots,K$, whereas
the synchronous instability mode corresponds to $\kap_1$ with eigenvector
$\omega_1=(1,\ldots,1)^T$.

Then, we use the explicit formulae for $c_j$ in (\ref{5s:mat_21}) and
for $v_0$ in (\ref{5:sum_1}) to write $\kap_i$ for $i=1,2$ as in
(\ref{5:nlep_2}). In addition, the ratio ${c_4/c_2}$ in
(\ref{5:vnlep}) can be calculated using (\ref{5s:mat_21}) and
(\ref{5:sum_1}) to get ${c_{4}/c_2}={2\tau_0\lam\beta/3}$, where
$\beta$ is defined in (\ref{5:nlep_3}). Finally, by diagonalizing the
vector NLEP (\ref{5:vnlep}) by using the matrix decomposition of
${\cal M}$, and by recalling the definition of ${\cal F}_i$ for
$i=0,1$ in (\ref{5s:f}), we obtain the NLEP (\ref{5:nlep}) of
Principal Result 5.2. This completes the derivation of Principal
Result 5.2 $\blacksquare$.

\setcounter{equation}{0}
\setcounter{section}{4}
\section{Discussion}
\label{sec:discuss}

\begin{figure}[tb]%
\[
\includegraphics[width=0.8\textwidth]{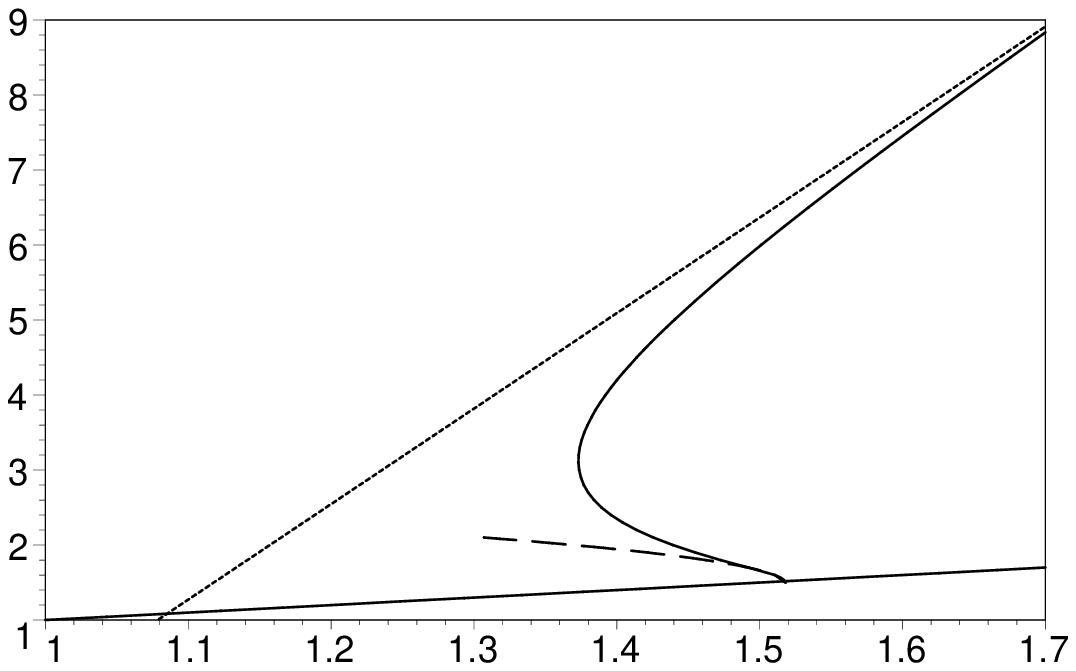}
\]

\setlength{\unitlength}{1\textwidth} \begin{picture}(0.98,0)(0,0)
\put(0.79, 0.29){\includegraphics[width=0.15\textwidth]{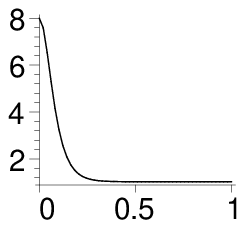}}
\put(0.64, 0.22){\includegraphics[width=0.15\textwidth]{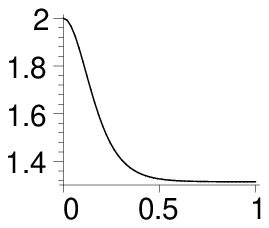}}
\put(0.7, 0.08){\includegraphics[width=0.15\textwidth]{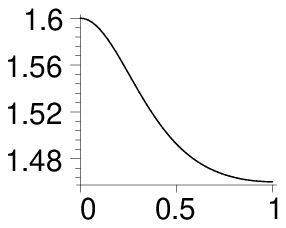}}
\put(0.4, 0.12){\includegraphics[width=0.15\textwidth]{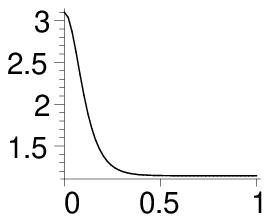}}
\put(0.7, 0.145){\vector(-1, -1){0.028}}
\put(0.66, 0.23){\vector(-1, -3){0.034}}
\put(0.45, 0.2){\vector(1,0){0.064}}
\put(0.825, 0.39){\vector(0,1){0.08}}
\put(0.5, 0){$\gamma$}
\put(0.03, 0.4){$A(0)$}
\end{picture}
\caption{Numerically computed bifurcation diagram of $A(0)$
  vs.~$\gamma$.  The parameter values are $\alpha=1, \varepsilon=0.05,
  x\in[0,1]$, and $D=2.$ A localized hot-spot appears for large values
  of $A(0)$. The asymptotics $A(0)\sim \frac{2(\gamma - \alpha)}{\eps
    \pi}$ (see (\ref{Aunif})) are shown by a dotted line. The constant
  steady state $A\sim\gamma$ is indicated by a solid straight line
  line. Turing patterns are born from the spatially uniform steady
  state as a result of a Turing bifurcation at $\gamma\sim
  {3\alpha/2}=1.5$. The weakly nonlinear regime is indicated by a
  dashed parabola coming out of the bifurcation point. Inserts shows
  the change in the shape of the profile $A(x)$ along the bifurcation
  curve. }%
\label{fig:bif}
\end{figure}

We have studied localized hot-spot solutions of (\ref{1:main}) in one
and two spatial dimensions in the regime $\eps^2\ll 1$ with $D\gg
1$. In this large $D$ limit, steady-state multi hot-spot solutions
have been constructed and their stability properties investigated with
respect to $D$ and $\tau$ from the analysis of certain nonlocal
eigenvalue problems.  An open problem is to characterize the dynamics
of multi hot-spot patterns by reducing (\ref{1:main}) to a finite
dimensional dynamical system for the locations of the hot-spots in a
quasi-steady-state pattern as was done for various two-component
reaction-diffusion models without drift terms in
\cite{dk,dkpr,iw,swr,kww_2,cw}.

We now remark on how the localized states constructed in this paper
are related to the weakly nonlinear Turing patterns studied in
\cite{s_1, s_2, s_3}.  In contrast to the theory developed in
\cite{s_2, s_3}, our parameter values for (\ref{1:main}) are not
restricted to lie close to the Turing bifurcation point.  In the limit
$\varepsilon\rightarrow0,$ the spatially homogeneous steady-state
solution for (\ref{1:main}) is $A_e=\gamma$ and
$P_e={(\gamma-\alpha)/\gamma}$. For $\eps\to 0$, it is linearly
unstable when (see equation (2.8) of \cite{s_2})
\begin{equation*}
\gamma>\frac{3}{2}\alpha \,, \qquad \mbox{as} \qquad \varepsilon\rightarrow
 0  \,,
\end{equation*}
and a spatially heterogeneous solution bifurcates off the homogeneous
steady state at $\gamma\sim\frac{3}{2}\alpha$ as
$\varepsilon\rightarrow0$. In contrast, a localized hot-spot
exists for the wider parameter range $\gamma>\alpha.$ In
Fig.~\ref{fig:bif} we plot the numerically computed bifurcation
diagram for $A(0)$ vs.~$\gamma$ on a one-dimensional interval, with
other parameters as indicated in the figure caption. The Turing
bifurcation at $\gamma=\frac{3}{2}\alpha$ is subcritical when $\eps\ll
1$ (cf.~\cite{s_2}), which is consistent with the stability of the
constant state when $\gamma<\frac{3}{2}\alpha$. However, the
bifurcation curve quickly turns around as it enters the localized
regime.  This is consistent with the existence of localized states
when $\gamma>\alpha.$

The dispersion relation obtained by linearizing (\ref{1:main}) about the
spatially uniform state $A_e$, $P_e$ is calculated as
\begin{equation}
\tau \lambda^2 + \lambda \left( A_e + Dm^2 + \tau (1-P_e) + 
 \tau \eps^2m^2 \right) + A_e (1+\eps m^2)- 3 D P_e m^2 + D m^2 (1+\eps^2 m^2)=0
 \,, \label{d:t}
\end{equation}
where $A_e=\gamma$ and $P_e={(\gamma-\alpha)/\gamma}$.  For
$\gamma>\frac{3}{2}\alpha$, it is readily shown from this relation
that the edges of the Turing instability band 
$m_{\text{lower}}<m<m_{\text{upper}}$ satisfy
\[
m_{\text{lower}}\sim D^{-1/2}\gamma(2\gamma-3\alpha)^{-1/2}\,; \qquad
 m_{\text{upper}}\sim\varepsilon^{-1}\gamma^{-1/2}(2\gamma
-3\alpha)^{1/2} \,, \qquad \text{as }\,\, \varepsilon\rightarrow0 \,.
\]
The most unstable mode (i.e. the one which grows the fastest, and
therefore the one most commonly observed) is obtained by setting
${d\lambda/dm}=0$ in (\ref{d:t}). For $\eps\to 0$, this gives the 
maximum growth rate
\begin{equation*}
  \lambda_{\text{dominant}} \sim 3 P_e -1 - 2\eps^2 m^2 \,,
\end{equation*}
together with the most unstable mode
\begin{equation*}
m_{\text{dominant}}\sim\varepsilon^{-1/2}D^{-1/4}\gamma^{-1/2}\left[
(\gamma-\alpha)(3\gamma^{2}+2\tau(2\gamma-3\alpha)\right]^{1/4}\,,
\end{equation*}
which is consistent with \cite{s_1} in the limit $\eps\to 0$ (see also
equation (2.9) of \cite{s_2}).  

Correspondingly, this implies that for an initial condition consisting
of a random perturbation of the spatially uniform steady-state, the 
preferred pattern has a characteristic half-length 
$l_{\text{turing}}\sim {\pi/m_{\text{dominant}}}$, where
\[
l_{\text{turing}}\sim\varepsilon^{1/2}D^{1/4}\gamma^{1/2}\left[
(\gamma-\alpha)(3\gamma^{2}+2\tau(2\gamma-3\alpha))\right]^{-1/4}\pi \,.
\]
In contrast, for localized structures the characteristic length 
$l$ between hot-spots to ensure stability of a multi hot-spot pattern, as
obtained from (\ref{Kc}), is that $l>l_c$ where
\begin{equation}
l_{c} \sim \sqrt{\pi}D^{1/4}\varepsilon^{1/2}\alpha^{1/2}(\gamma-\alpha)^{-3/4}
\,.
\end{equation}
Although $l_{\text{turing}}$ and $l_{c}$ are not related, they are
both of the same asymptotic order $O(D^{1/4}\varepsilon^{1/2})$, which
implies that the number of stable localized hot-spots corresponds roughly
to the most unstable Turing mode. In fact, for a large parameter
range, the inequality $\l_c<l_{\text{turing}}$ holds. For example,
when $\tau=1, \alpha=1, \gamma=2$ we calculate that
$l_c/l_{\text{turing}}=0.7<1$. This was already observed empirically
in Fig.~6 of \cite{s_1}.

\begin{figure}[tb]
\[
\includegraphics[width=0.95\textwidth]{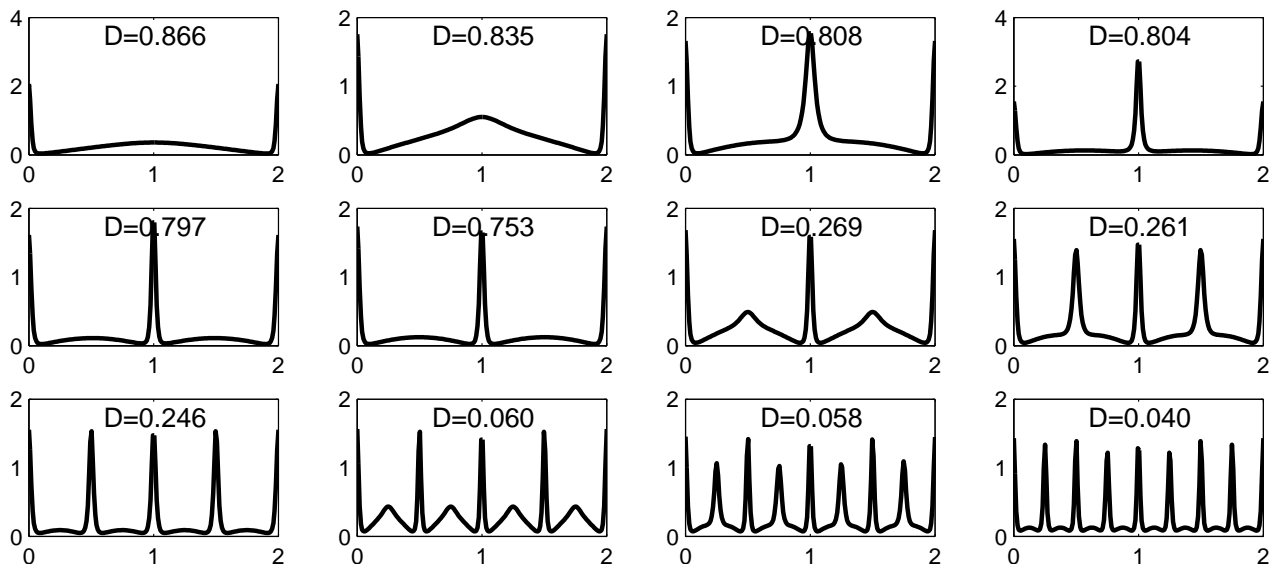}
\]
\caption{Hot-spot insertion phenomenon. Numerical solution of 
  (\ref{1:main}) with $\eps=0.02, \alpha=1, \gamma=2, D=1/(1+0.01 t)$
  with $x \in (0,2).$ Initial conditions consist of two boundary
  hot-spots. Snapshots of $P(x)$ are shown for values of $D$ as
  indicated. Hot-spot insertion takes place every time that $D$ is
  quartered.}
\label{fig:insert}%
\end{figure}

Formula (\ref{Kc}) provides an upper bound $K_{c}$ on the number $K$
of stable hot-spots in the regime where $D\gg 1$. On the other hand,
numerical evidence shows the existence of a {\em lower bound} that
occurs when $D$ is sufficiently small. In this regime an
instability occurs when there are \emph{too few} hot-spots.  In fact,
if the hot-spot inter-distance exceeds some critical length, then a
hot-spot insertion phenomena is observed (see Fig.~\ref{fig:insert}).
From Fig.~\ref{fig:insert}, it appears that hot-spot insertion takes
place every time that $D$ is quartered.  A similar insertion
phenomenon occurs in other reaction-diffusion systems such as a
chemotaxis model \cite{painter-hillen}, the Gray-Scott model
\cite{pearson}, the ferrocyanide-iodide-sulfite system
\cite{lee-swinney}, the Brusselator \cite{kww-bru}, the Schnakenburg
model \cite{kww_2}, and a model of droplet breakup \cite{liu-bert}.
The detailed analysis of hot-spot insertion phenomena in
(\ref{1:main}) will be considered in future work.

\section*{Acknowledgements} 
T.~K.~and M.~J.~W.~were supported by NSERC Discovery Grants (Canada). 
J.~W.~was supported from an Earmarked Grant of the RGC of Hong Kong.


\begin{thebibliography}{99}

\bibitem{cw} W.~Chen and M.~J.~Ward (2009),
  \textit{Oscillatory instabilities and dynamics of multi-spike
    patterns for the one-dimensional Gray-Scott model},
  Europ. J. Appl.  Math {\bf 20}(2), pp.~187--214.

\bibitem{cw_1} W.~Chen and M.~J.~Ward (2011),
\textit{The stability and dynamics of localized spot patterns in the 
two-dimensional Gray-Scott model}, SIAM J. Appl. Dyn. Sys., {\bf 10}(2),
(2011), pp.~582--666.

\bibitem{dgk_0} A.~Doelman, R.~A.~Gardner and T.~J.~Kaper (2001), {\em Large
stable pulse solutions in reaction-diffusion equations}, Indiana U. Math.
J., {\bf 50}(1), pp.~443--507.

\bibitem{dgk_1} A.~Doelman, R.~A.~Gardner and T.~J.~Kaper (2002), {\em A
stability index analysis of 1-D patterns of the Gray Scott model},
Memoirs of the AMS, {\bf 155}, No.~737.

\bibitem{dgk_2} A.~Doelman, R.~A.~Gardner and T.~J.~Kaper
(1998), \textit{Stability analysis of singular patterns in the
1D Gray-Scott model: A matched asymptotic approach}, Physica D, {\bf
122}(1-4), pp.~1--36.

\bibitem{dk} A.~Doelman and T.~J.~Kaper (2003), \textit{Semistrong
pulse interactions in a class of coupled reaction-diffusion systems},
SIAM J. Appl. Dyn. Sys., {\bf 2}(1), pp.~53--96.

\bibitem{dkpr} A.~Doelman, T.~J.~Kaper and K.~Promislow (2007),
\textit{Nonlinear asymptotic stability of the semi-strong pulse
dynamics in a regularized Gierer-Meinhardt model}, SIAM
J. Math. Anal., {\bf 38}(6), pp.~1760--1789.

\bibitem{HP} T.~Hillen and A.~Potapov (2004), \textit{The one-dimensional
 chemotaxis model: global existence and asymptotic profile}, Math. Meth. 
 Appl. Sci., {\bf 27}, pp.~1783--1801.

\bibitem{iw} D.~Iron and M.~J.~Ward (2002), \textit{The dynamics of
multi-spike solutions to the one-dimensional Gierer-Meinhardt model},
SIAM J. Appl. Math., {\bf 62}(6), pp.~1924--1951.

\bibitem{iww} D.~Iron, M.~J.~Ward and J.~Wei (2001), \textit{The
stability of spike solutions to the one-dimensional Gierer-Meinhardt
model}, Physica D, {\bf 150}(1-2), pp.~25--62.

\bibitem{iweiwin} D.~Iron, J.~Wei and M.~Winter (2004), \textit{Stability
  analysis of Turing patterns generated by the Schnakenberg model},
  J. Math. Biol., {\bf 49}(4), pp. 358--390.

\bibitem{kkw_0} K.~Kang, T.~Kolokolnikov and M.~J.~Ward (2007), 
\textit{The stability and dynamics of a spike in a one-dimensional
Keller-Segel model}, IMA J. Appl. Math., {\bf 72}(2), pp.~140--162.

\bibitem{kmjw} T.~Kolokolnikov and M.~J.~Ward (2003),
\textit{Reduced-wave Green's functions and their effect on the
dynamics of a spike for the Gierer-Meinhardt model},
Europ. J. Appl. Math., {\bf 14}(5), pp.~513--545.

\bibitem{kshad} T.~Kolokolnikov and M.~J.~Ward (2004), \textit{Bifurcation of
  spike equilibria in a near shadow reaction-diffusion System}, DCDS-B,
  {\bf 4}(4), pp.~1033--1064.

\bibitem{kww_1} T.~Kolokolnikov, M.~J.~Ward and J.~Wei (2005),
\textit{The existence and stability of spike equilibria in the
one-dimensional Gray-Scott model: The low feed-rate regime}, Studies
in Appl. Math., {\bf 115}(1), pp.~21--71.

\bibitem{kww_2} T.~Kolokolnikov, M.~J.~Ward and J.~Wei
(2009), \textit{Spot self-replication and dynamics for the
Schnakenburg model in a two-dimensional domain}, J. Nonlinear Sci., {\bf
19}(1), pp.~1--56.

\bibitem {kww-bru}T.~Kolokolnikov, M.~J.~Ward, and J.~Wei (2007),
\textit{Self-replication of mesa patterns in reaction-diffusion models}, 
Physica D, {\bf 236}(2), pp.~104--122.

\bibitem{kww_3} T.~Koloklonikov, M.~J.~Ward and J. Wei (2006),
  \textit{Slow translational instabilities of spike patterns in the
   one-dimensional Gray-Scott model}, Interfaces and Free Boundaries,
    {\bf 8}(2), pp.~185--222.

\bibitem{kw} T.~Kolokolnikov and J.~Wei (2011), \textit{Stability of
spiky solutions in a competition model with cross-diffusion},
SIAM J. Appl. Math. \textbf{71}, pp.~1428--1457.

\bibitem{lee-swinney}K.~J.~Lee and H.~L.~Swinney (1995),
  \textit{Lamellar structures and self-replicating spots in a
    reaction-diffusion systems}, Phys. Rev. E., \textbf{51}(3),
  pp.~1899--1915.

\bibitem{liu-bert} W. Liu, A. L. Bertozzi, and T. Kolokolnikov (2012),
  \textit{Diffuse interface surface tension models in an expanding
    flow}, Comm. Math. Sci., \textbf{10}(1), pp.~387--418.

\bibitem{mckay} R.~McKay and T.~Kolokolnikov (2012), \textit{Stability
transitions and dynamics of localized patterns near the shadow limit of
reaction-diffusion systems}, to appear, DCDS-B.

\bibitem{mo_2} C.~B.~Muratov and V.~V.~Osipov (2002),
  \textit{Stability of static spike autosolitons in the Gray-Scott
    model}, SIAM J. Appl. Math., {\bf 62}(5), pp.~1463--1487.

\bibitem{mo_3} C.~B.~Muratov and V.~V.~Osipov
(2000), \textit{Static spike autosolitons in the Gray-Scott model},
J. Phys. A: Math Gen.,  {\bf 33}, pp.~8893--8916.

\bibitem{n} Y.~Nishiura (2002), \textit{Far-from equilibrium
dynamics, translations of mathematical monographs}, Vol.~{\bf 209},
AMS Publications, Providence, Rhode Island.

\bibitem{painter-hillen} K. Painter and T. Hillen (2011), {\emph
  Spatio-temporal chaos in a chemotaxis Model}, Physica D, {\bf 240},
  pp.~363--375.

\bibitem {pearson} J.~E.~Pearson (1993), \textit{Complex Patterns in a
  Simple System}, Science, {\bf 216}, pp.~189--192.

\bibitem{PH} A.~Potapov and T.~Hillen (2005), \textit{Metastability in 
chemotaxis models}, J. Dynam. Diff. Eq., {\bf 17}(2) pp.~293-330.

\bibitem{s_1} M.~B.~Short, M.~R.~D'Orsogna, V.~B.~Pasour, G.~E.~Tita,
  P.~J.~Brantingham, A.~L.~Bertozzi and L.~B.~Chayes (2008), \textit{A
    statistical model of criminal behavior},
  Math. Models. Meth. Appl. Sci., {\bf 18}, Suppl.  pp.~1249--1267.

\bibitem{s_2} M.~B.~Short, A.~L.~Bertozzi and P.~J.~Brantingham (2010),
\textit{Nonlinear patterns in urban crime - hotpsots, bifurcations, and
suppression}, SIAM J. Appl. Dyn. Sys., {\bf 9}(2), pp.~462--483.

\bibitem{s_3} M.~B.~Short, P.~J.~Brantingham, A.~L.~Bertozzi and G.~E.~Tita
(2010), \textit{Dissipation and displacement of hotpsots in reaction-diffusion
models of crime}, Proc. Nat. Acad. Sci. {\bf 107}(9) pp.~3961-3965.

\bibitem{sww} B.~Sleeman, M.~J.~Ward and J.~Wei (2005), \textit{The
existence and stability of spike patterns in a chemotaxis model},
SIAM J. Appl. Math., {\bf 65}(3), pp.~790--817.

\bibitem{swr} W.~Sun, M.~J.~Ward and R.~Russell (2005),
\textit{The slow dynamics of two-spike solutions for the Gray-Scott
and Gierer-Meinhardt systems: competition and oscillatory
instabilities}, SIAM J. Appl. Dyn. Syst., {\bf 4}(4), pp.~904--953.

\bibitem{vpd} H.~Van der Ploeg and A.~Doelman (2005), \textit{Stability of
spatially periodic pulse patterns in a class of singularly perturbed
reaction-diffusion equations}, Indiana Univ. Math. J., {\bf 54}(5),
 p.~1219-1301.

\bibitem{mjww_1} M.~J.~Ward and J.~Wei (2003), \textit{Hopf
bifurcations and oscillatory instabilities of spike solutions for the
one-dimensional Gierer-Meinhardt model}, J. Nonlinear Sci., {\bf
13}(2), pp.~209--264.

\bibitem{mjww_2} M.~J.~Ward and J. Wei (2002), \textit{The existence and 
 stability of asymmetric spike patterns in the Schnakenburg model}, Studies 
 in Appl. Math., {\bf 109}(3), pp.~229--264.

\bibitem{mjww_3} M.~J.~Ward and J. Wei (2002), \textit{Asymmetric spike
   patterns for the one-dimensional Gierer-Meinhardt model: equilibria
   and stability}, Europ. J. Appl. Math., {\bf 13}(3), (2002),
    pp.~283--320.

\bibitem{ww_shad} M.~J.~Ward and J.~Wei (2003), \textit{Hopf bifurcation
of spike solutions for the shadow Gierer-Meinhardt model}, Europ. J. Appl.
Math., {\bf 14}(6), pp.~677--711.

\bibitem{ww_1} J.~Wei and M.~Winter (2001), \textit{Spikes for the
two-dimensional Gierer-Meinhardt system: the weak coupling case}, J.
Nonlinear Sci., {\bf 11}(6), pp.~415--458.

\bibitem{ww_6} J.~Wei and M.~Winter (2002), \textit{Spikes for the
two-dimensional Gierer-Meinhardt system: the strong coupling case}, J.
Diff. Eq., {\bf 178}, pp.~478--518.

\bibitem{ww_2} J.~Wei and M.~Winter (2003),
\textit{Existence and stability of multiple spot solutions for the
Gray-Scott model in $\mathbb{R}^2$}, Physica D., {\bf 176}(3-4), pp.~147-180.

\bibitem{ww_3} J.~Wei and M.~Winter (2008),
\textit{Stationary multiple spots for reaction-diffusion systems},
J. Math. Biol., {\bf 57}(1), pp.~53--89.

\bibitem{ww_4} J.~Wei and M.~Winter (2004), \textit{Existence and stability
analysis of asymmetric patterns for the Gierer-Meinhardt system},
J. Math. Pures Appl. (9), {\bf 83}(4), pp.~433-476.

\bibitem{ww_5} J.~Wei and M.~Winter (2003), \textit{Asymmetric spotty
patterns for the Gray-Scott model in } $\R^2$, Studies in Appl. Math.,
{\bf 110}(1), pp.~63--102.

\bibitem{zhang} J.~Wei and L.~Zhang (1998), \textit{On a nonlocal eigenvalue
 problem}, Ann. Sc. Norm. Sup. Pisa C1. Sci. pp.~41--62.

\bibitem{wei_rev} J.~Wei (2008), \textit{Existence and stability of
  spikes for the Gierer-Meinhardt system}, book chapter in
  \textit{Handbook of Differential Equations, Stationary Partial
    Differential Equations}, Vol. 5 (M. Chipot ed.), Elsevier,
  pp.~489--581.

\end{thebibliography}
\end{document}